\documentclass[preprint2,times]{aastex62}
\usepackage{apjfonts}
\usepackage{enumitem}
\usepackage{color}

\newcommand{\eqref}[1]{Equation (\ref{#1})}

\newcommand{\figref}[1]{Figure~\ref{#1}}

\newcommand\Tstrut{\rule{0pt}{2.6ex}}

\shorttitle{Sub-Chandrasekhar Detonations}

\begin{document}

\title{Quantifying How Density Gradients and Front Curvature\\ Affect Carbon Detonation Strength During Type~Ia Supernovae}

\author{Broxton J. Miles}
\affiliation{Department of Physics, North Carolina State University, Raleigh, NC, USA, Email:bjmiles@ncsu.edu}
\affiliation{Department of Physics \& Astronomy,
		University of Alabama, Tuscaloosa, AL, USA}
\author{Dean M. Townsley}
\affiliation{Department of Physics \& Astronomy,
	University of Alabama, Tuscaloosa, AL, USA}
\author{Ken. J Shen}
\affiliation{Department of Astronomy and Theoretical Astrophysics Center,
	University of California, Berkeley, CA, USA}
\author[0000-0002-0474-159X]{F.~X.~Timmes}
\affiliation{School of Earth and Space Exploration, Arizona State University, Tempe, AZ, USA}
\affiliation{Joint Institute for Nuclear Astrophysics - Center for the Evolution of the Elements, USA}
\author{Kevin Moore}
\affiliation{Independent Researcher, email: kevinelmoore@gmail.com}

\begin{abstract}
Accurately reproducing the physics behind the detonations of Type Ia supernovae and the resultant nucleosynthetic yields is important for interpreting observations of  spectra and remnants.
The scales of the processes involved span orders of magnitudes, making the problem computationally impossible to ever fully resolve in full star simulations in the present and near future.
In the lower density regions of the star, the curvature of the detonation front will slow the detonation, affecting the production of intermediate mass elements.
We find that shock strengthening due to the density gradient present in the outer layers of the progenitor is essential for understanding the nucleosynthesis there, with burning extending well below the density at which a steady-state detonation is extinct.
We show that a complete reaction network is not sufficient to obtain physical detonations at high densities and modest resolution due to numerical mixing at the unresolved reaction front.
At low densities, below 6$\times$10$^{5}$~g~cm$^{-3}$, it is possible to achieve high enough resolution to separate the shock and the reaction region,and the abundance structure predicted by fully resolved quasi-steady-state calculations is obtained.
For our best current benchmark yields, we utilize a method in which the unresolved portion of Lagrangian histories are reconstructed based on fully resolved quasi-steady-state detonation calculations.
These computations demonstrate that under-resolved simulations agree approximately, $\sim$10\% in post-shock values of temperature, pressure, density, and abundances, with expected detonation structures sufficiently far from the under-resolved region, but that there is still room for some improvement in the treatment of subgrid reactions in the hydrodynamics to before better than 1$\%$ can be achieved at all densities.

\end{abstract}


\section{Introduction}
Type Ia supernovae are generally agreed upon to be the result of the thermonuclear disruption of carbon-oxygen white dwarfs. 
Apart from this, there remains little that can be said with certainty about the nature of these events.
What is known is that these events are extremely luminous, powered by the radioactive decay of $^{56}$Ni \citep{pankey,Colgate69}, and when combined with a well constrained relationship between light curve decline time and peak luminosity, type Ia supernovae can be used as standard candles \citep{phillips.lira.ea:reddening-free}.
This feature has allowed SNe Ia to be used to measure redshifts out to $z\sim1$, making SNe Ia instrumental in the discovery of the  acceleration of the expansion of the universe and by consequence, the existence of dark energy \citep{riess.filippenko.ea:observational,perlmutter.aldering.ea:measurements}.
Much of the uncertainty around SNe Ia lies in the nature of the progenitor system.
Progenitor systems of SNe Ia fall into two categories based on the number of white dwarfs involved: single degenerate or double degenerate.
From these categories, the exploding white dwarf can be placed into two mass categories, Chandrasekhar mass or sub-Chandrasekhar mass.
The single degenerate, Chandrasekhar mass scenario \citep{Whelan73} is the canonical picture of a white dwarf accreting material until it crosses the Chandrasekhar mass limit of $\sim$1.4~M$_\odot$, whereupon nuclear reactions begin and eventually disrupt the white dwarf.
In the sub-Chandrasekhar scenario, also known as the double detonation scenario \citep{Nomo82_1}, the progenitor white dwarf accretes helium either peacefully from a non-degenerate companion or through a violent merger with a white dwarf companion.
When enough has been accreted, a detonation is triggered in the helium shell and in the proper conditions may trigger the detonation of the carbon-oxygen core of the white dwarf.
A recent study by \citet{shen_d6} has found evidence that three high velocity, galactic white dwarfs may be the ejected companion stars from this scenario.

\begin{figure}
	\includegraphics[width=\columnwidth]{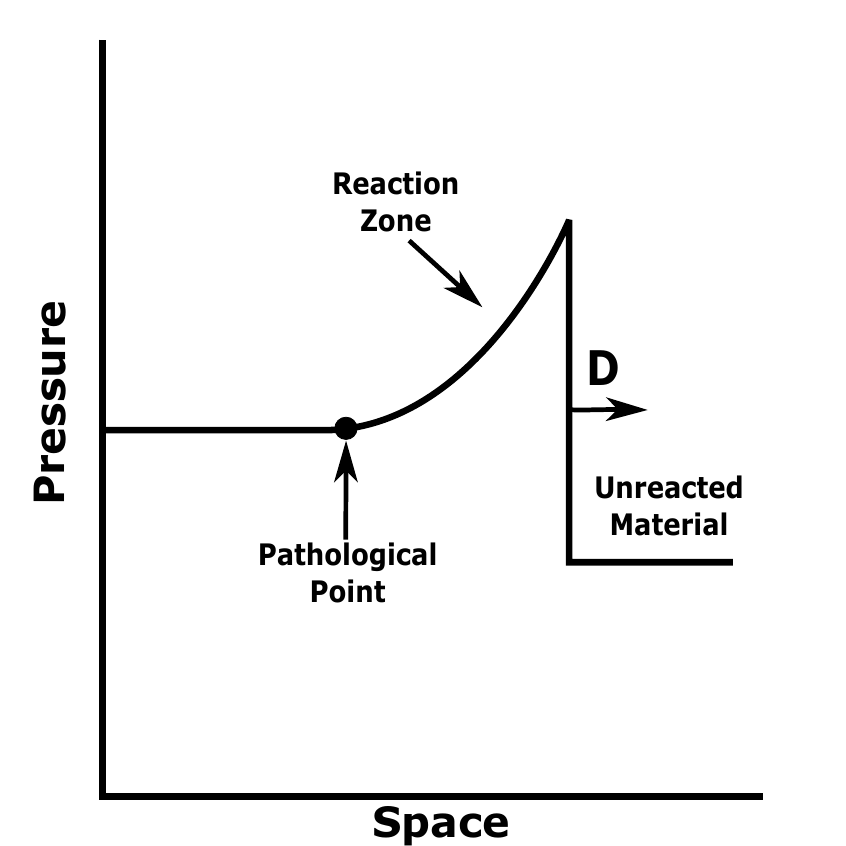}
	\caption{\label{znd_schem}
	Schematic of a ZND, eigenvalue, planar detonation adapted from \cite{carvallo_2012}.}
\end{figure}

In the case of a sub-Chandrasekhar mass progenitor or, for a Chandrasekhar mass progenitor during the second stage of a delayed detonation, the fusion reactions that power the destruction of the star propagate supersonically as detonations.
The simplest 1D model of a detonation that agrees fairly well with reality is known as the Zeldovich-von Neumann-Doering (ZND) model \citep{zeldovich_1940,vonneumann_42,vonneumann_63,doring_43} and is shown in figure \ref{znd_schem}.
In the ZND model, a planar shock wave moves into an unreacted medium with some speed, D.
The material is shocked to a pressure and density state based on the von Neumann conditions.
Once the material reaches this condition, reactions occur in a region behind the shock known the reaction zone.
The energy produced in this region is deposited into the shock, powering its forward propagation.
At some point, the sonic point, the flow of material behind the shock will become super-sonic with respect to and causally disconnected from the shock.
For Chapman-Jouguet detonations, the sonic point occurs at the termination of the reaction zone.
If one of the reactions behind the shock front is endothermic or there is some other form of dissipation, the sonic point moves into the reaction zone, and the Chapman-Jouguet detonation is no longer possible.
In this case, the lowest detonation speed that provides a well behaved solution to the governing equations is the eigenvalue speed, and the produced detonation is known as the eigenvalue or pathological detonation. \citep{Fickett,sharpe_1999,bdzil_review}

\begin{figure}
	\includegraphics[width=\columnwidth]{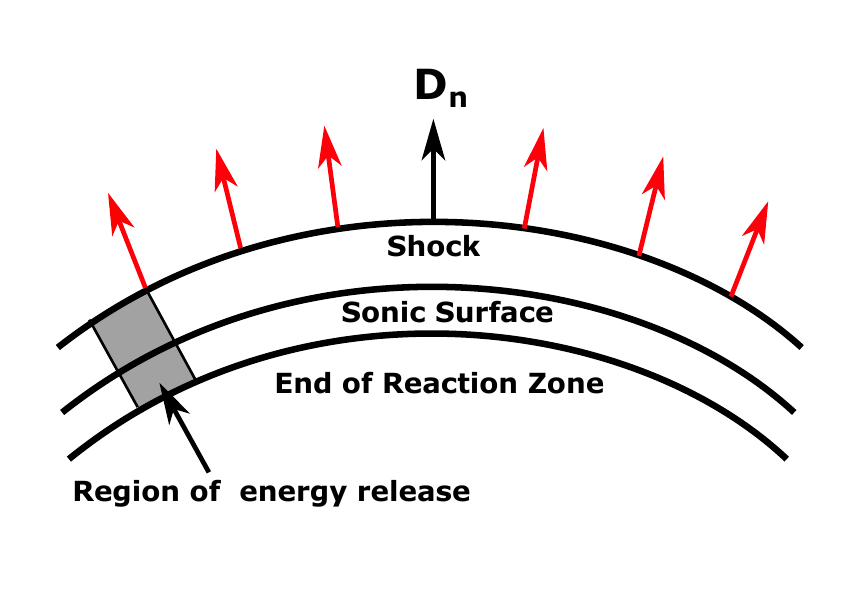}
	\caption{\label{curved_det}
		Schematic of a detonation with positive curvature, adapted from \cite{bdzil_review}. The speed of a positively curved detonation is lower for two reasons. First, curvature introduces divergence to the flow (red arrows) that reduces the energy available to the forward propagation of the detonation. Second, curvature moves the location of the sonic surface inward, reducing the area of the reaction zone that is causally connected to the shock.}
\end{figure}

\begin{figure}
	\includegraphics[width=\columnwidth]{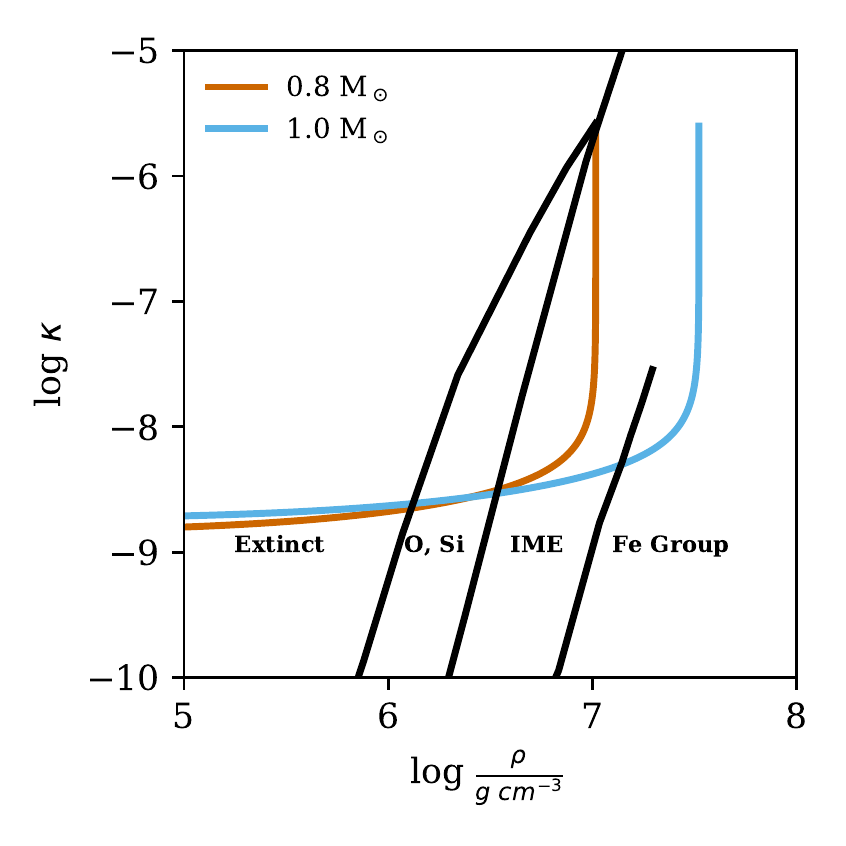}
	\caption{\label{curv_fig}
		Detonation curvature vs density. 
		Curvature, $\kappa$ is measured as 1/radius of curvature. In this 1-D scenario, the radius of curvature is the radial location in the progenitor.
		The orange and blue lines show the measured curvature for a 0.8 M$_\odot$ and 1.0 M$_\odot$ progenitor, respectively.
		The black solid lines \citep{Dunkley_et_al_2013} separate regions of the expected nucleosynthetic yields from an eigenvalue detonation at a given density and curvature.
		Below $\sim$2$\times$10$^{6}$ g cm$^{-3}$, the detonation is expected to become extinct, and no longer self-propagating. }
\end{figure}

As a consequence of the star's geometry, detonations powering SNe Ia will not be planar, but curved.
\figref{curved_det} gives a simple schematic of a curved detonation.
In this 2-D diagram, the sonic point has become a sonic surface, but overall the structure is similar to the planar case.
A detonation which is convex in the direction of propagation will be weaker than the planar case for two main reasons.
First, curvature introduces flows that are divergent from the direction of propagation.
Divergent flows will remove energy that would otherwise be used to power the forward propagation of the shock.
Secondly, the sonic surface is moved inward reducing the area of the causally connected region of the reaction zone, thereby reducing the energy available to power the forward propagation of the shock.
Taking into account the detonation strength is  crucial due to its determination of nucleosynthetic products.
\cite{townsley_moore} and \cite{Moore_et_al_2013}  demonstrated the importance of curvature for detonations in helium shells on white dwarfs.
\cite{Dunkley_et_al_2013} studied the effects of detonation curvature on the nucleosynthetic products of carbon burning from SNe Ia.
They found that in the mid to low density regions of the white dwarf, small amounts of curvature are sufficient to change the final products of the detonation.
Their findings are shown as the black lines in figure \ref{curv_fig}.
This figure shows the expected nucleosynthetic products given an eigenvalue detonation initiated at a selected density and curvature, $\kappa$.
$\kappa$ is calculated as 1/radius of curvature, and in 1-D, is 1/radial coordinate within the white dwarf. 
The red line shows the values of $\kappa$ for the progenitor used in this study.
From this figure, we can expect that there should be a large amount of intermediate mass and lighter elements produced in the explosion.
At $\sim$2.0$\times 10^{6}$ g~cm$^{-3}$, the detonation is expected to become extinct, and no further burning should occur. 

\begin{figure}
	\includegraphics[width=\columnwidth]{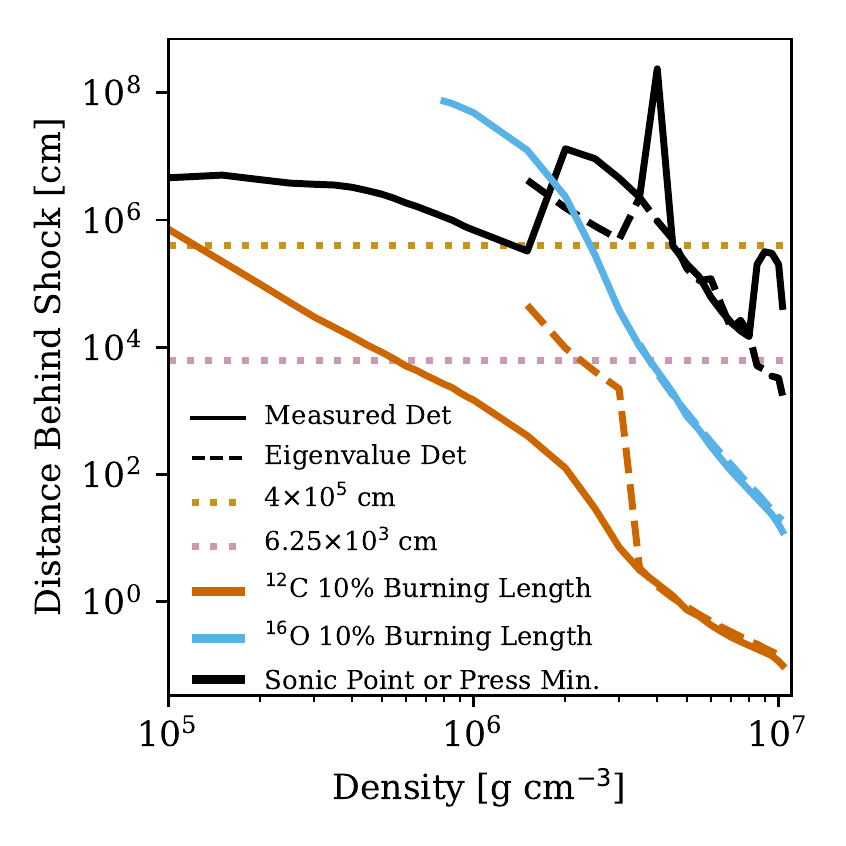}
	\caption{\label{burning_length_fig}
	The distance behind the shock at which 10\% of $^{12}$C (orange) and $^{16}$O (blue) has been burned for a steady-state, eigenvalue detonation (dashed) as well as a steady-state detonation traveling at speeds taken from the 0.0625 km resolution calculation (solid).
	The curvature for each density is taken from the 0.8 M$_\odot$ progenitor.
	The purple dotted line represents the highest resolution from the explosion simulation, and the gold dotted curve line represents a resolution typically used in type Ia explosion simulations.
	Burning for each stage is unresolved for densities at which the solid line falls below the dotted line.
	Also shown in black is the location of the sonic point (for eigenvalue detonations) or for overdriven detonations, the location of the analogous pressure minimum.}
\end{figure}

The main problem associated with accurately modeling detonations in full-star calculations of SN Ia comes from the length scales involved.
Figure \ref{burning_length_fig} shows the distance behind the shock at which 10$\%$ of the peak values of carbon (orange) and oxygen (blue) have been consumed for a steady state detonation at the given density and its appropriate curvature from figure \ref{curv_fig}.
The solid line is a similar length scale but for the higher detonation speeds observed in the simulation due to the shock steepening induced by the density gradient in the outer layers of the progenitor (see section~\ref{sec:progen}).
The purple, dotted line represents the highest resolution calculation we have completed, 0.0625 km, and the  brown, dotted line represents a typical resolution used in full-star type Ia calculations, 4 km 
The portion of the blue and orange lines that falls under each dotted line represents unresolved burning.
At the highest densities, the carbon consumption length reaches the order of centimeters and smaller, making it, essentially, computationally impossible to fully resolve the entire explosion in full star simulations.
The unresolved nature of the detonations raises questions about the fidelity of the results that are produced by hydrodynamic simulations (see section \ref{hydro_results}).
In the past, the solution to this problem has been to post-process temperature-density histories recorded during the simulation by tracer particles.
While this may produce results that are more valid than those from the hydro calculations, one should realize that the unresolved features of the detonation are imprinted on the temperature-density histories as well \citep{Harris_2017}.
Consequently, directly post-processing the temperature-density histories still leaves the thermodynamic structure of the detonation under-resolved.

This paper is novel in three ways.
First, we demonstrate a new method of reconstructing the unresolved portion of the detonation during post-processing.
Second, we assess the important aspects that can influence a detonation, including density dependence, curvature, and density gradients over a wide range of spatial resolutions.
Third, we verify benchmark yields by comparing to the structure of steady-state detonations as given by the ZND model computed with a fully resolved error-controlled integration and appropriately complete nuclear network.
Reconstructing the unresolved detonation structures in the post-processing gives us confidence in our ability to produce accurate, verified benchmark yields that can be used in comparisons to observations, as well as the results from other simulations. 
As our benchmark problem we choose a centrally ignited sub-Chandrasekhar full detonation.
This demonstrates all the important aspects that can influence detonations, including density dependence, curvature, and density gradients, in a configuration that can be simulated in one dimension, allowing a wide range of resolutions.
Over a small range of low densities, we are even able to resolve the reaction front in the hydrodynamics.
We choose to focus on a 0.8 M$_\odot$ progenitor since it will have a smaller portion of its mass burned completely to Fe-group elements compared to higher mass progenitors, making computation of accurate yields most challenging.

In section~\ref{sec:progen}, we give detailed description of the progenitor and the explosion simulation and its results. 
We then discuss our methods of post-processing and give a detailed description of the reconstruction method in section~\ref{sec:post_proc}.
Results of the explosion simulation, direct post-processing are discussed in section~\ref{sec:results} and are compared to the post-processed results of other other calculations that utilized alpha-particle networks, parameterized burning schemes, and thickened detonation fronts in section~\ref{sec:aprox13}.

\section{Progenitor and Software} 
\label{sec:progen}
In this work, we use a 0.8 M$_\odot$  white dwarf with initial mass fractions of 50 percent carbon, 48.6 percent oxygen, and 1.4 percent metals corresponding to solar metal mass fractions \cite{Asplund_metals}, as the progenitor.
The central density and temperature of the progenitor was 1.05$\times 10^7$ g~cm$^{-3}$ and 3$\times 10^7$ K, respectively.
The detonation of the progenitor was initiated by placing a 150 km temperature gradient with a peak temperature of 1.98$\times 10^9$ K into the core of white dwarf, with the required size of the gradient determined by calculations similar to those in \cite{Seitetal09}. 

\begin{figure}
	\includegraphics[width=\columnwidth]{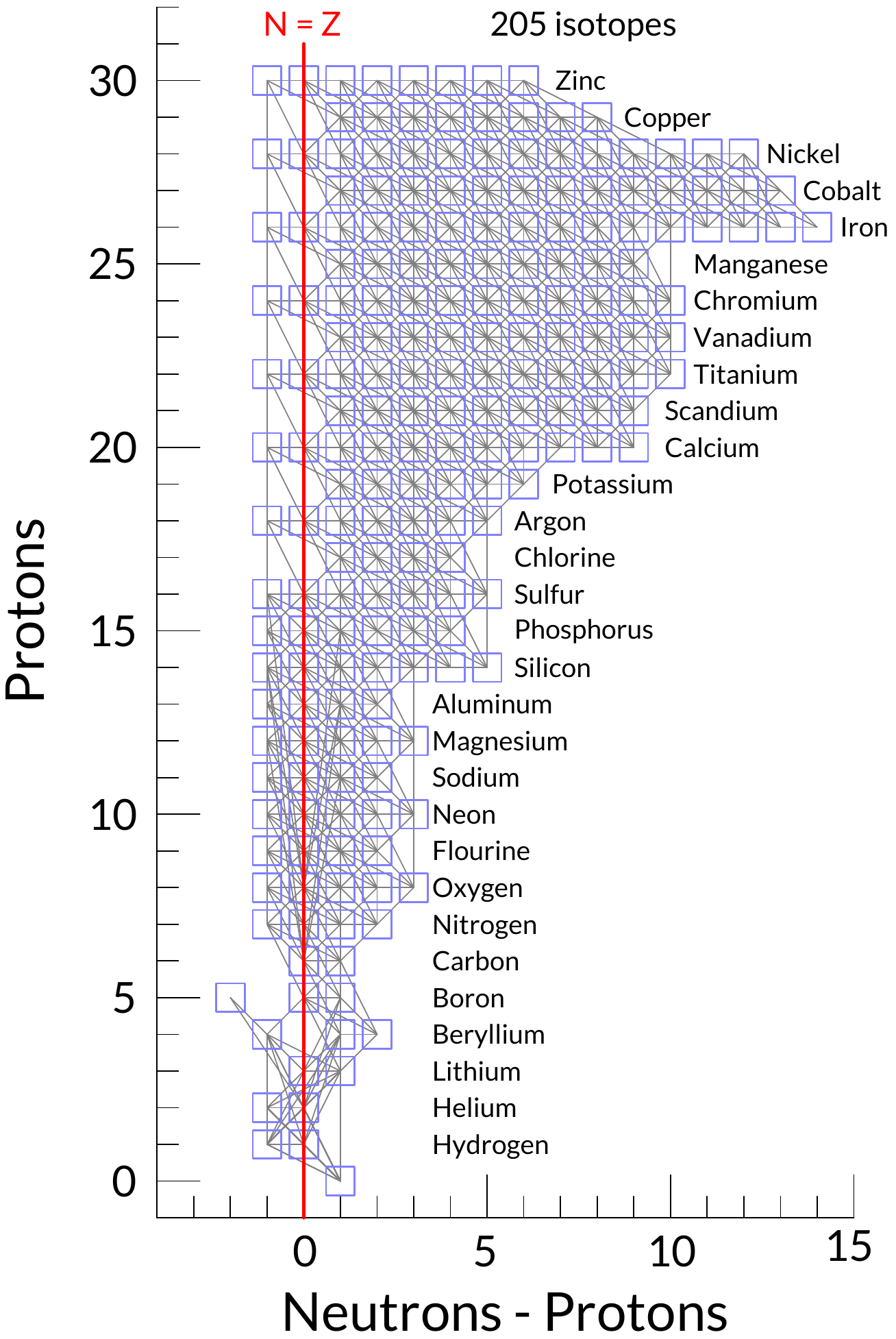}
	\caption{\label{fig:network_schematic} Schematic of the 205 isotope network used in the explosion simulations and post-processing calculations.}
\end{figure}

We utilize the FLASH hydrodynamics code \citep{fryxell_2000_aa} for the explosion simulation segment of this study.
The explosions simulations were run in one dimension at 4, 2, 1, 0.5, 0.25, 0.125, and 0.0625 km resolution for 7.5 seconds of evolution.
To minimize computational cost, an adaptive refinement technique is used.
The simulation is only fully refined in regions where the energy generated by nuclear reactions is higher than 10$^{16}$~ergs~g$^{-1}$~s$^{-1}$.
For regions which do not meet this energy generation criteria, the adaptive mesh is managed as described in \cite{Townsleyetal09}.
At 0.8 seconds, the shock has left the star and there is no longer a reaction front.
During the last 6.7 seconds of simulation the time, the star is allowed to derefine, reducing the cost of the simulation, but not significantly affecting the overall results from each calculation.
As previously performed in \cite{Shen_18} (Shen18) and used here for the first time with a fairly complete reaction network, we have integrated the nuclear reaction network from the Modules for Experiments in Stellar Astrophysics \citep[MESA,][]{paxton_2011,paxton_2013,paxton_2015,paxton_2018} into FLASH.
This allows for the use of arbitrary sized networks during the explosion simulation, contrary to our previous studies, in which a small network or parameterized burner is used during the explosion simulation, and a larger network is used to post-process the results. 
In these simulations, we use a 205-isotope network that consists of neutrons, $^{1-2}$H,$^{3-4}$He, $^{6-7}$Li, $^{7,9-10}$Be, $^{8,10-11}$B, $^{12-13}$C, $^{13-16}$N,$^{15-19}$O,$^{17-20}$F, $^{19-23}$Ne, $^{21-24}$Na, $^{23-27}$Mg, $^{25-28}$Al, $^{27-33}$Si, $^{29-34}$P, $^{31-37}$S, $^{35-38}$Cl, $^{35-41}$Ar, $^{39-44}$,K, $^{39-49}$Ca, $^{43-51}$Sc, $^{43-54}$Ti, $^{47-56}$V, $^{47-58}$Cr, $^{51-59}$Mn, $^{51-66}$,Fe, $^{55-67}$Co, $^{55-68}$Ni, $^{59-66}$Cu, and $^{59-66}$Zn.
Figure~\ref{fig:network_schematic} shows a schematic of this network.
We use the nuclear reaction network from MESA release 7624 with the latest JINA-reaclib reaction rates that are contained within that release.
For comparison purposes, explosion simulations were also completed using the aprox13 \citep{Timmes_aprox13} nuclear reaction network and our parameterized burning routine described in \cite{townsley_2016}.

As will be discussed in section~\ref{recon}, we use the open source software instrument eZND \citep{Moore_et_al_2013}, to reconstruct the detonation in our post-processing.
eZND is a tool that allows us to calculate the results of steady-state detonations in the ZND paradigm for a given pre-detonation density, background temperature, radius of curvature, and detonation speed. 
We chose to use eZND for two primary reasons.
One, it utilizes the same MESA nuclear reaction network that is used in both our explosion simulations and direct post-processing so we can use an identical nuclear reaction network across all three methods.
Two, compared to our previous work \citep[][Townsley16]{townsley_2016}, which only implements planar detonations, eZND includes the effects of curvature during the ZND integration \citep{sharpe_2001} and includes methods for automated determination of the eigenvalue detonation as well as traversal of the pathological point \citep{sharpe_1999}, and has been used in the study of helium shell detonations \citep{Shen_14,Moore_et_al_2013}
\section{Explosion Simulation Results}
\label{hydro_results}
\begin{deluxetable*}{c|ccccccc}
\tablewidth{\columnwidth}
\tablecaption{Selected Explosion Simulation Yields by Resolution in M$_\odot$ \label{tab:hydro_yields}}
\tablehead{%
	\colhead{Iso} & \colhead{4km} & \colhead{2km} & \colhead{1km} & \colhead{0.5km} & \colhead{0.25km} & \colhead{0.125km} & \colhead{0.0625km}
}
\startdata
$^{28}$Si & 0.188 & 0.206 & 0.223 & 0.239 & 0.256 & 0.271 & 0.285 \\
\hline
$^{32}$S\phm{.}\Tstrut & 0.130 & 0.138 & 0.145 & 0.150 & 0.154 & 0.158 & 0.161 \\
\hline
$^{40}$Ca\Tstrut & 0.0323 & 0.0321 & 0.0317 & 0.0313 & 0.0303 & 0.0291 & 0.0276 \\
\hline
$^{56}$Ni\Tstrut & 0.194 & 0.170 & 0.147 & 0.124 & 0.103 & 0.0843 & 0.068 
\enddata
\end{deluxetable*}
\begin{figure*}\centering{
	\includegraphics[]{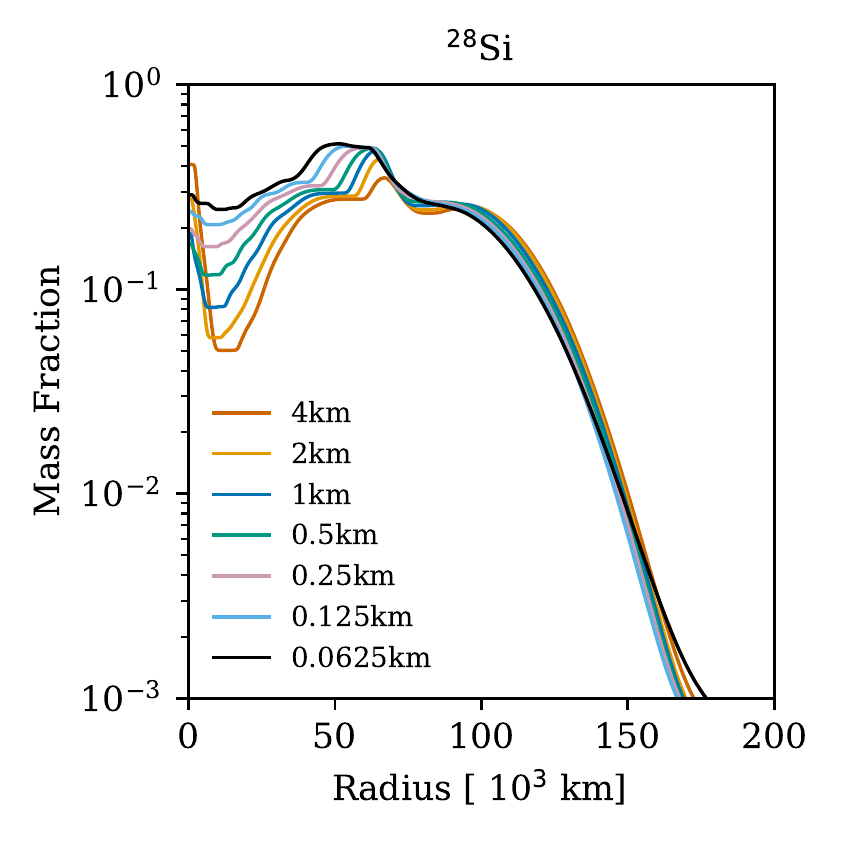}\
	\includegraphics[]{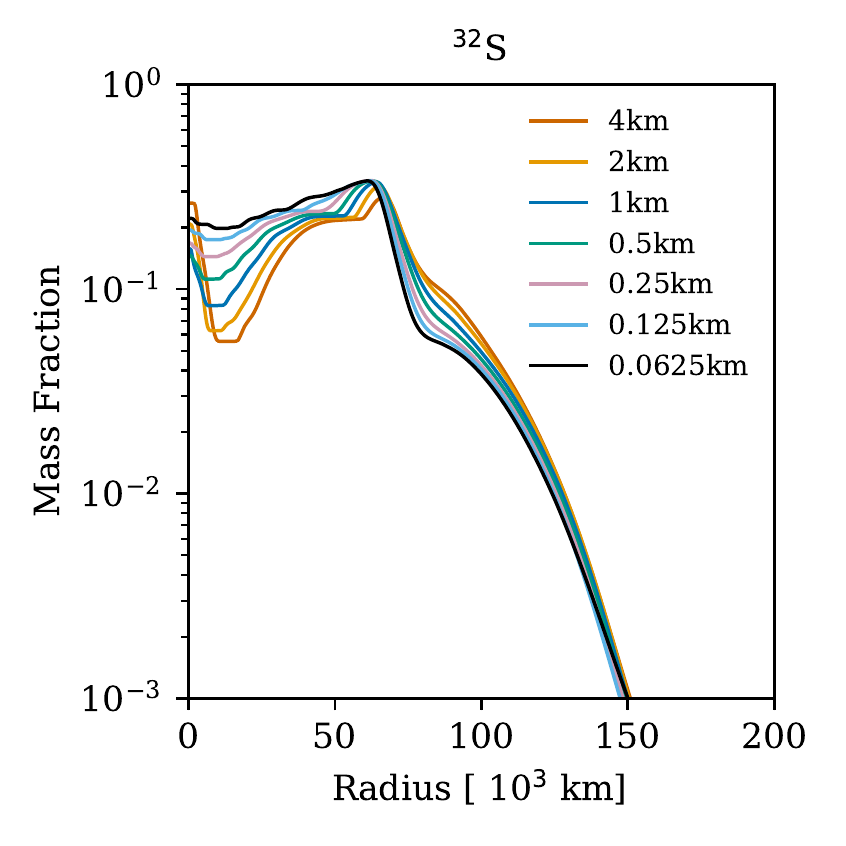}
	\includegraphics[]{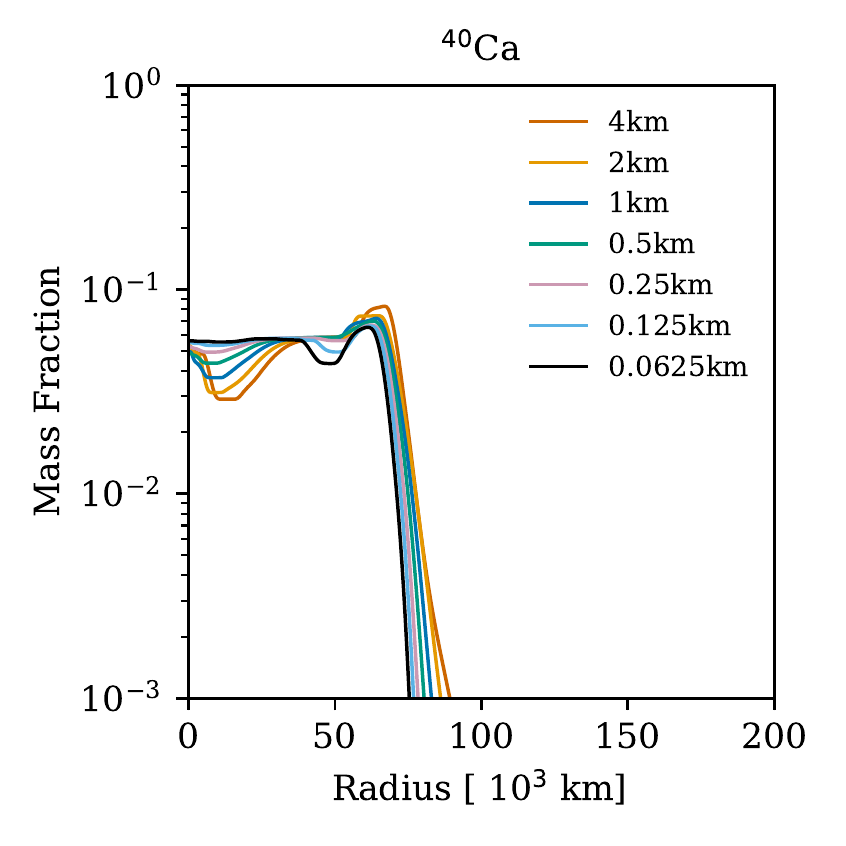}\	
	\includegraphics[]{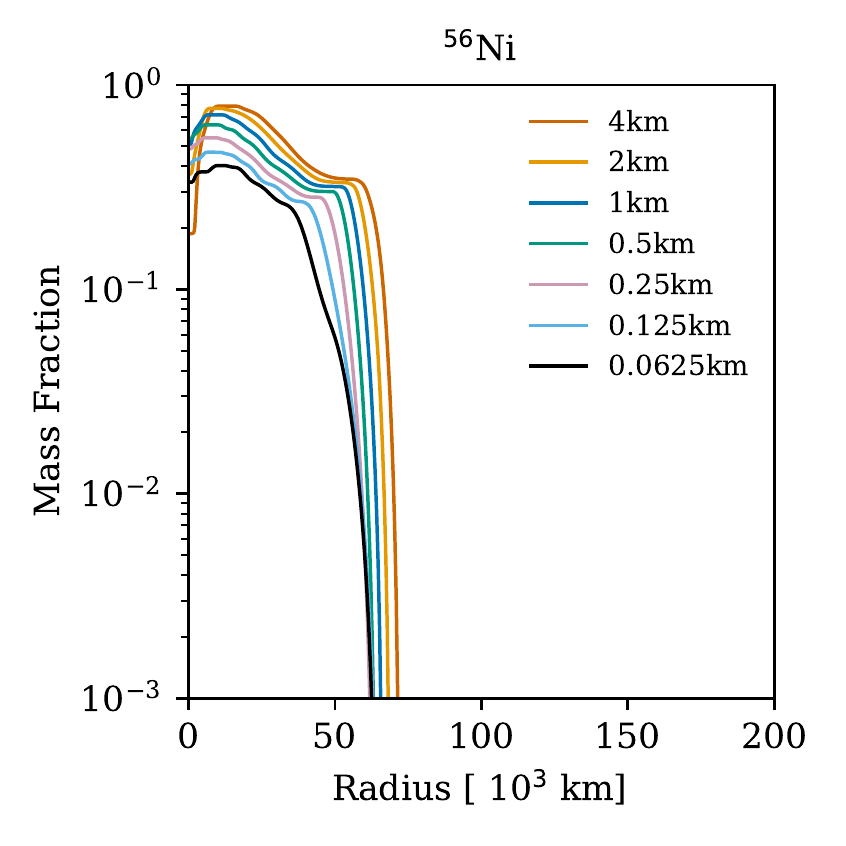}}
	\caption{\label{hydro_rad_profile}
	Isotopic mass fractions of $^{28}$Si (top,left), $^{32}$S (top, right), $^{40}$Ca (bottom,left), and $^{56}$Ni (bottom,right) vs radius from the explosion simulations run at 7.5 seconds post-explosion at 4 (red), 2 (orange), 1 (blue), 0.5 (green), 0.25 (magenta), 0.125 (cyan), and 0.0625 (black) km resolutions. The mass fraction profiles all show a strong dependence on the resolution, though the direction and strength depend on the location in star. 
	}
\end{figure*}
\begin{figure}
	\includegraphics[width=\columnwidth]{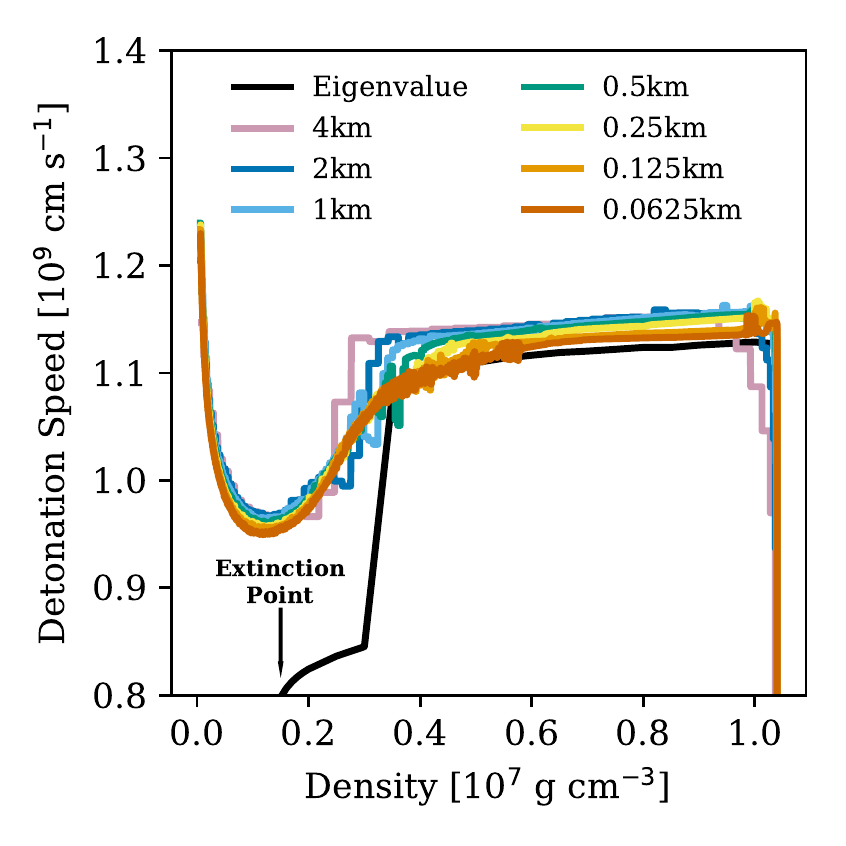}
	\caption{\label{det_speed_fig}
	Detonation speeds measured from the explosion simulations and for a steady-state (eigenvalue) detonation are shown for various densities
	The speeds in higher resolution simulations are closer to the expected eigenvalue speed except at lower densities, where the shock is significantly stronger than that of a steady-state detonation due to the density gradient present in the progenitor.
	}
\end{figure}
\begin{figure*}\centering{
	\includegraphics{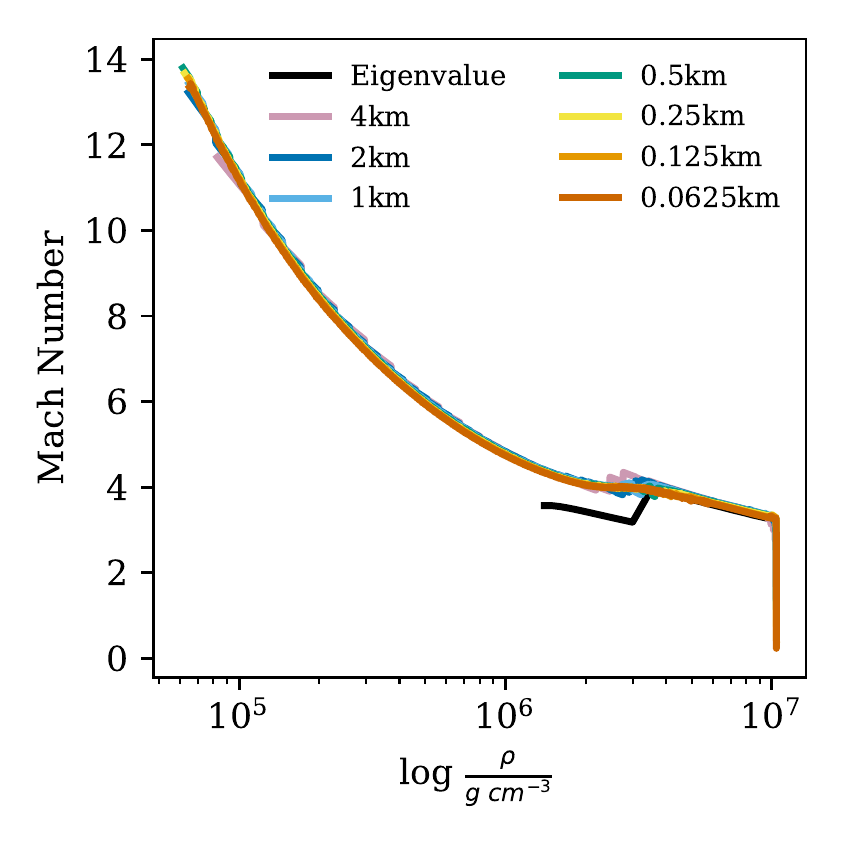}
	\includegraphics{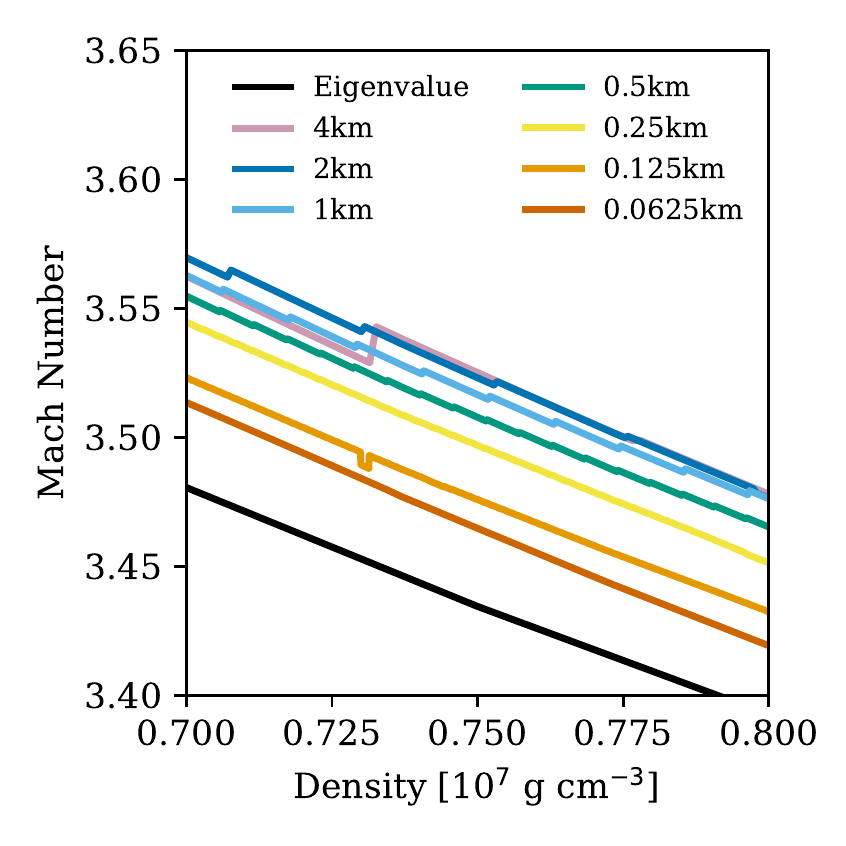}}
\caption{\label{fig:mach}Mach number as a function of density. The detonations from the explosion simulations have consistently higher Mach numbers than the eigenvalue detonation. The Mach number of the explosion simulation detonation approaches the Mach number of the eigenvalue detonation with increasing resolution. }
\end{figure*}

In this section, we discuss the results of the explosion simulations, and show that due to the unresolved nature of the detonation, we must use post-processing in some form to obtain reliable nucleosynthetic results.
\figref{hydro_rad_profile} shows the isotopic mass fraction profiles of four of the major nucleosynthetic products of our explosion simulations at 7.5 seconds post-explosion, $^{28}$Si, $^{32}$S, $^{40}$Ca, and $^{56}$Ni, at 4, 2, 1, 0.5, 0.25, 0.125, 0.0625 km resolution.
For each of the shown isotopic mass fractions, there is a clear dependence on the resolution of the simulation, though the strength and direction vary.
The profiles of $^{32}$S and $^{40}$Ca are both of importance due to their use in comparisons to observations and $^{56}$Ni is the energy source of the observed light from SNe Ia.
Early-time spectral features are dependent on the abundances in the outer layers of the material ejected by the explosion.
At radii greater than $\sim$75000 km, the mass fraction of $^{32}$S can vary by as much as a factor of 2, and the mass fraction of $^{40}$Ca can vary by orders of magnitude as the resolution increases.
Abundance variations of this scale would likely have a non-negligible effect on observed spectral features.

The produced mass fractions of $^{56}$Ni do not fare much better with increasing resolution.
Since it is produced in the densest region of the star, the burning lengths are at their shortest, and this is burning that can not be fully resolved.
Although the yields are suspect, the amount of $^{56}$Ni produced gives us insight into the strength of the detonation as it travels through the white dwarf.
As the resolution increases, the peak mass fraction of $^{56}$Ni falls, pointing to the decreasing strength of the detonation with increasing resolution.

Table~\ref{tab:hydro_yields} contains the yields of $^{12}$C, $^{16}$O, $^{28}$Si, $^{32}$S, $^{40}$Ca, and $^{56}$Ni in M$_\odot$ taken from the explosion simulations at each resolution.
Most striking is the change in the $^{56}$Ni yield from 4~km to 0.0625~km resolution.
The amount of $^{56}$Ni produced falls from 0.194 M$_\odot$ to 0.068 M$_\odot$, a factor of $\approx$2.9. 
As seen in the radial profiles, as the resolution increases the yields of the products of complete and incomplete silicon burning decrease and the silicon yield increases, again adding evidence for a weaker explosion at higher resolution.

To confirm the resolution's effect on the detonation's speed and thus strength, we also measured the velocity of the detonation in the explosion simulation.
The position of the detonation was recorded at each timestep in the simulation.
A line was fit to the position-time data in 100 timestep intervals and the slope of the line was taken as the detonation speed.
Figure \ref{det_speed_fig} shows the calculated detonation speeds as a function of density.
As the resolution increases, the measured speed of the detonation decreases, pointing to a weaker detonation, agreeing with the nucleosynthetic results.
Though the nucleosynthetic results of our explosion simulations are not converged at this point, the convergence of the measured detonation speeds is much better, appearing essentially converged over a wide range of densities.
Also shown in this figure is the calculated eigenvalue detonation speed at a given density with the curvature corresponding to the WD radius at the location of the detonation.
As expected from figure \ref{curv_fig}, the eigenvalue detonation speed decreases before hitting the transition from the primarily incomplete Si burning branch to the primarily carbon-oxygen burning branch, and then finally becoming extinct.
The detonation in the explosion simulation travels faster than the eigenvalue detonation at all resolutions, though the detonation in the 0.0625 km case briefly dips to the eigenvalue speed at $\sim$5$\times 10^6$ g~cm$^{-3}$.
This is due to the extra energy released by the simultaneous burning of carbon and oxygen (section~\ref{sec:results}, figure~\ref{fig:abund_example}).
Where it most differs is near the extinction point.
Here the eigenvalue detonation has become extinct due to the combination of the low density and the curvature, but the detonation in the explosion simulation begins to accelerate. 
At this location in the progenitor there is a strong density gradient.
Far from decreasing strength below 3$\times$10$^6$ g cm$^{-3}$ and going extinct below 2$\times$10$^6$ g cm$^{-3}$ as was assumed by \cite{Dunkley_et_al_2013}, we find that the detonation is strongly overdriven when traveling down the density gradient.

Figure~\ref{fig:mach} shows the Mach number of the detonation.
The Mach number provides another way of measuring the strength of the detonation.
Across all models, the explosion simulation detonations have consistently higher Mach numbers than the eigenvalue detonation.
The Mach numbers of the explosion simulations systematically decrease toward the Mach number of the eigenvalue detonation with increasing resolution.

While the detonation speed in our simulation appears fairly converged, the difference between this and the eigenvalue speed demonstrates that an under-resolved calculation can appear converged and still not be accurate.
The region of figure~\ref{det_speed_fig} where we expect the eigenvalue solution to be most accurate is at densities in the range of 0.7$\times$10$^7$~g~cm$^{-3}$. 
This is the region of the star where the density gradient is most shallow and the detonation is most thin, leading to what should be a very close match to the eigenvalue solution.
We attribute the remaining differences between the detonation speeds in this region to the under-resolved calculation.
As seen in figure~\ref{burning_length_fig}, the $^{16}$O burning length scale in this region is between 10 and 100 cm, and the $^{12}$C burning length scale is a few tenths of a cm.
These are so far below the grid scale that in order to do better it will be necessary to improve the treatment of sub-grid-scale reactions in the simulation, most likely with some form of direct model.
Section~\ref{sec:shen} discusses the use of a reaction limiter, which should perform better but is short of a direct model.
However, the improvement is found to be modest.
We consider this to be a remaining source of uncertainty in our benchmark yields.

The detonation speed observed in the simulation is higher than the eigenvalue solution at all densities, most significantly at lower densities.
This corresponds to an overdriven state, mostly induced by the shock propagating down a density gradient.
While a steady-state overdriven solution contains no sonic point and has a density minimum \citep[see][for examples]{townsley_2016}, here we use it as an approximate proxy for the abundance structure near the shock as described in more detail in section~\ref{recon}.

The effects the strength of the detonation has on the burning lengths of carbon and oxygen at a given density are shown in figure~\ref{burning_length_fig}.
The solid lines are the burning lengths of carbon (orange) and oxygen (blue) and calculated from a steady-state detonation at a given density with the appropriate curvature and detonation speed measured from the 0.0625~km resolution explosion simulation, and the dashed lines are the burning lengths measured from an eigenvalue speed detonation at the same conditions.
In the case of carbon the overdriven detonation burning length is slightly shorter than the eigenvalue detonation at densities from $\sim$3.5$\times$10$^{6}$ g cm$^{-3}$ to $\sim$1.0$\times$10$^{7}$ g cm$^{-3}$.
At lower densities, when the eigenvalue detonation transitions from the incomplete Si burning branch, the overdriven detonation burning lengths are orders of magnitude shorter.
As mentioned above, the eigenvalue detonation reaches a density where it can no longer propagate, but the overdriven detonation continues.
At $\sim$6.0$\times$10$^{5}$ g cm$^{-3}$, the burning length becomes resolved in the 0.065 km calculation.
The oxygen burning length shows only a light change at densities higher than the extinction density.
However, much like the carbon burning, oxygen burning is allowed to continue at lower densities with the stronger, overdriven detonation.
Oxygen burning becomes resolved in the 0.0625 km calculation at $\sim$3.5$\times$10$^{6}$ g cm$^{-3}$.
Also shown in figure~\ref{burning_length_fig}, are the distances to either the sonic point, for eigenvalue detonations, or the pressure minimum, for overdriven detonations.
In general, as the density decreases, the location of the sonic point or pressure minimum moves further behind the shock front.
When the detonation reaches a density where curvature causes it to change burning branches, the distance to the sonic point or pressure minimum undergoes a sudden change.
At ~2$\times$10$^{6}$~g~cm$^{-3}$, the oxygen consumption length becomes longer than the distance to the sonic point or pressure minimum for both the detonation calculated at simulation conditions and the eigenvalue detonation.

The role of overdriving on the strength of the detonation means that techniques that use results from eigenvalue detonations are insufficient to describe the detonation behavior at these lower densities.
This suggests that consideration of the strengthening of the detonation due to density gradients in the star, which depends on the relative direction of the detonation propagation with respect to the density gradient in multi-dimensional simulations, is critical to understanding and therefore modeling and validating, nucleosynthetic yields for SNe Ia.

\section{Post-Processing}
\label{sec:post_proc}

In this section we will the discuss the two methods we use for post-processing the temperature-density histories produced in our explosion simulations.
We will first discuss the method of directly post-processing the histories using the MESA one-zone burner. 
We then go on to discuss our new method of reconstructing the detonation during post-processing.
\subsection{Direct Particle Post-Processing}
Each explosion simulation carried 10000 Lagrangian tracer particles that recorded temperature-density histories for post-processing.
At the end of the simulation, the particles were divided into 100 velocity bins in increments of 250 km/s.
This is similar to the process used in \cite{townsley_2016}, we have now utilized the MESA one-zone burner to integrate the temperature-density histories, using the same 205 isotope nuclear reaction network that was included in the explosion simulation.
The particles in each velocity bin are post-processed, and then the average of each bin is taken.
The ejecta density profile is obtained by binning the results of the final simulation timestep into 1000 radial bins.
The total mass, average density, and average radial velocity of material in each bin is computed. 
The velocity in each bin is then used to interpolate the isotopic mass fractions, which were computed in velocity bins, into mass bins. 
\subsection{Post-Processing with Detonation Reconstruction}
\label{recon}
As we have established, the structure of the detonation in the explosion simulation is unresolved through large portions of the star.
Direct post-processing may be enough to overcome this problem, but there still exists the problem that the unresolved portion of the detonation is imprinted in the temperature-density histories used as the input.
By reconstructing the detonation during post-processing we are able to compute yields with time histories in which the detonation physics is fully numerically resolved throughout.
The uncertainty in the derived nucleosynthesis is then related to accuracy of the instantaneous detonation speed, and therefore shock strength, computed in the simulation.
Relating uncertainty to a physical parameter creates a more manageable situation from the standpoint of systematic uncertainty than the ad-hoc use of numerically under-resolved histories in post-processing.

\begin{figure*}
	\centering\includegraphics[]{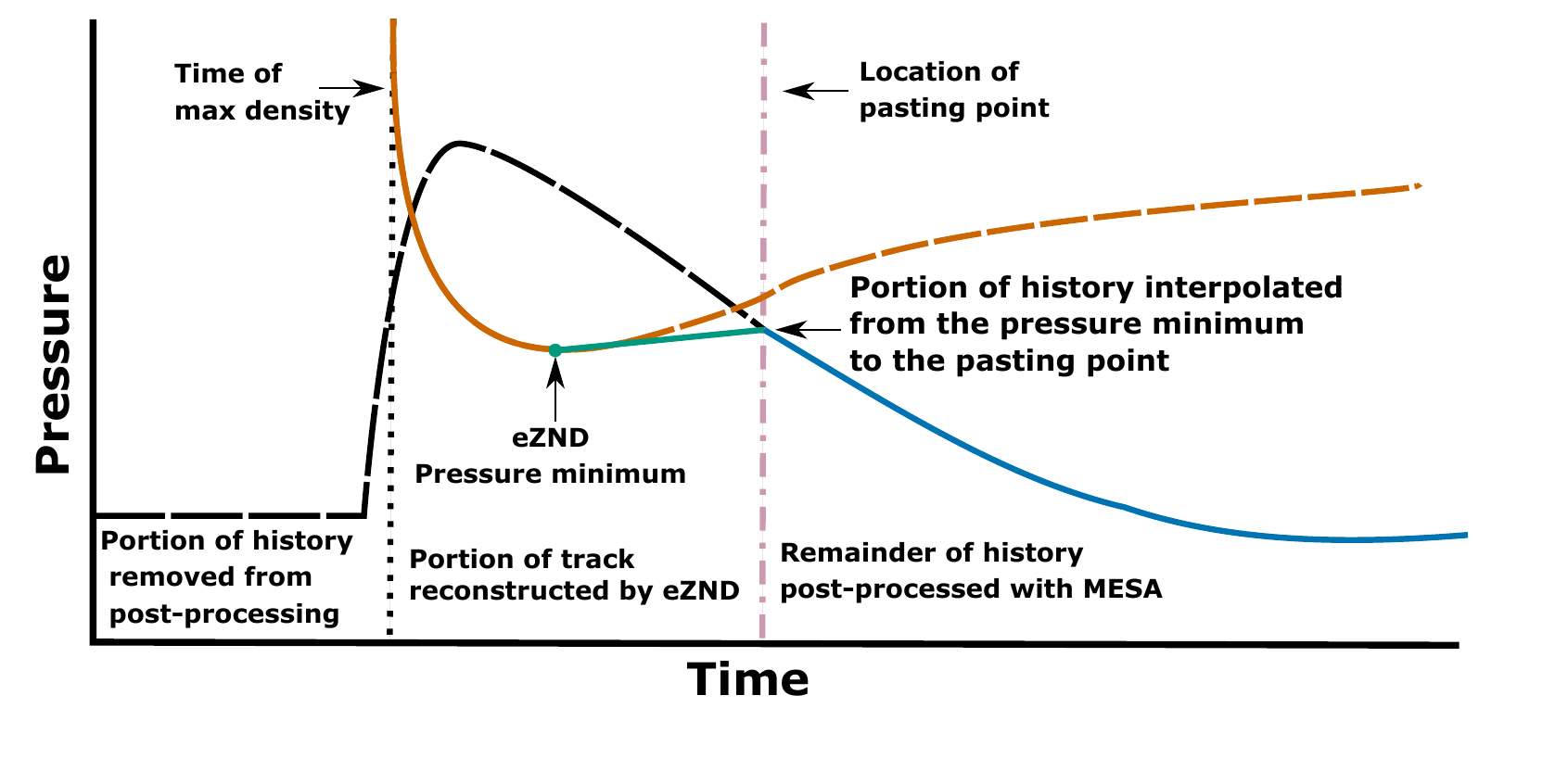}
	\centering\includegraphics{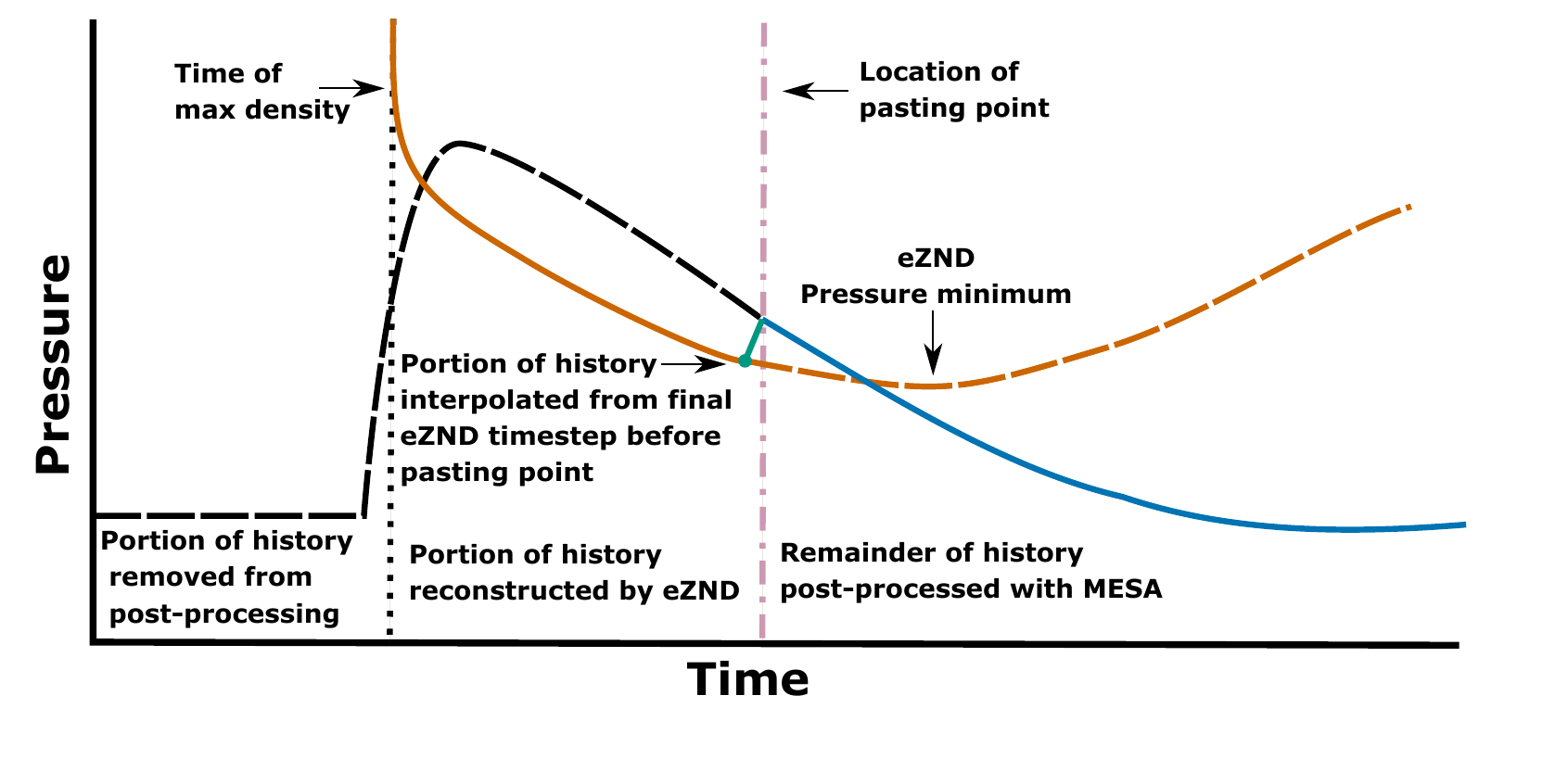}
	\centering\caption{\label{fig:recon_illustration} 
		Schematic of the reconstruction of an under-resolved temperature-density history shown in pressure. 
		The quasi-steady-state detonation calculation (solid, orange) replaces the unresolved detonation structure (dashed, black) recorded by the explosion simulation from the time of maximum density (dotted, black) to the chosen pasting point (dash-dotted, purple).
		The initial point of the final MESA integration is determined by the location of pasting point relative to the location the pressure minimum of the eZND integration.
		If the pasting point occurs after the pressure minimum (top), the MESA integration begins at the time of the pressure minimum and the intermediate history is interpolated between the pressure minimum and the pasting point (green) and then continues the integration using the remainder of the temperature-density history (blue).
		If the pasting point occurs before the pressure minimum (bottom), the initial point of the final integration begins at the final eZND integration timestep before the pasting point.}
	
\end{figure*}

Reconstruction is done for each post-processed track in the explosion simulation.
The pre-detonation density, background temperature, and radius of curvature can all be taken directly from a particle's starting position in the progenitor.
This leaves only the detonation speed as the tunable parameter.
We have chosen to use the measured detonation speed from the explosion simulation (figure \ref{det_speed_fig}) at a track's pre-detonation density as the detonation speed in its reconstruction.
Reconstructing the history of a track is a multistage process illustrated in figure~\ref{fig:recon_illustration}.
First, we must pick a starting time for the detonation.
We choose that to be the time of the density peak recorded by the track as it moves through the shock.
The pressure peak is another valid choice, but we choose the starting time to coincide with the time of the recorded maximum density (henceforth, density peak), as it consistently occurs the earliest in the under-resolved tracks.
Next, we choose a point, that will be referred to as the pasting point, some number of steps after the density peak.
In this work, we have chosen the pasting point to be 20 steps after the density peak.
Now, the history between the density peak and the pasting point is replaced by the results of a ZND integration of a steady-state detonation, performed with eZND, at the track's pre-detonation density, temperature, and radius of curvature with the supplied detonation speed up to the time of the pasting point.
Finally, the remainder of the track is post-processed using the MESA one-zone burner with the final isotopic mass fractions of the ZND integration used as the initial mass fractions.

The starting point of the final integration is determined by the pasting point's location relative to the pressure minimum in the ZND integration.
Figure~\ref{fig:recon_illustration} gives an illustration of the following scenarios.
If the pasting point occurs after the pressure minimum, the temperature and density at the pressure minimum of the ZND integration are used as the starting point, and the temperature and density history is interpolated to the pasting time.
Conversely, if the pasting point occurs before the pressure minimum, the integration begins at the final point of the eZND integration before the pasting point with the intermediate history interpolated between that point and the pasting point.

We use the term "quasi-steady-state" to refer to the detonation structure that results from the above pasting procedure.
It should be noted that the detonation speed taken from the explosion simulation is usually higher than the eigenvalue speed, so that that the portion of the history computed with eZND corresponds to an overdriven, or supported, steady-state solution. 
The overdriven steady-state solution is only an approximation of the structure of a detonation moving down a density gradient.
Evaluation of systemic uncertainty due to this approximation is left to future work, but is expected to be small based on the similarity of the tracks and the overdriven steady-state solution. 

\begin{figure*}
\centering\includegraphics[scale=1]{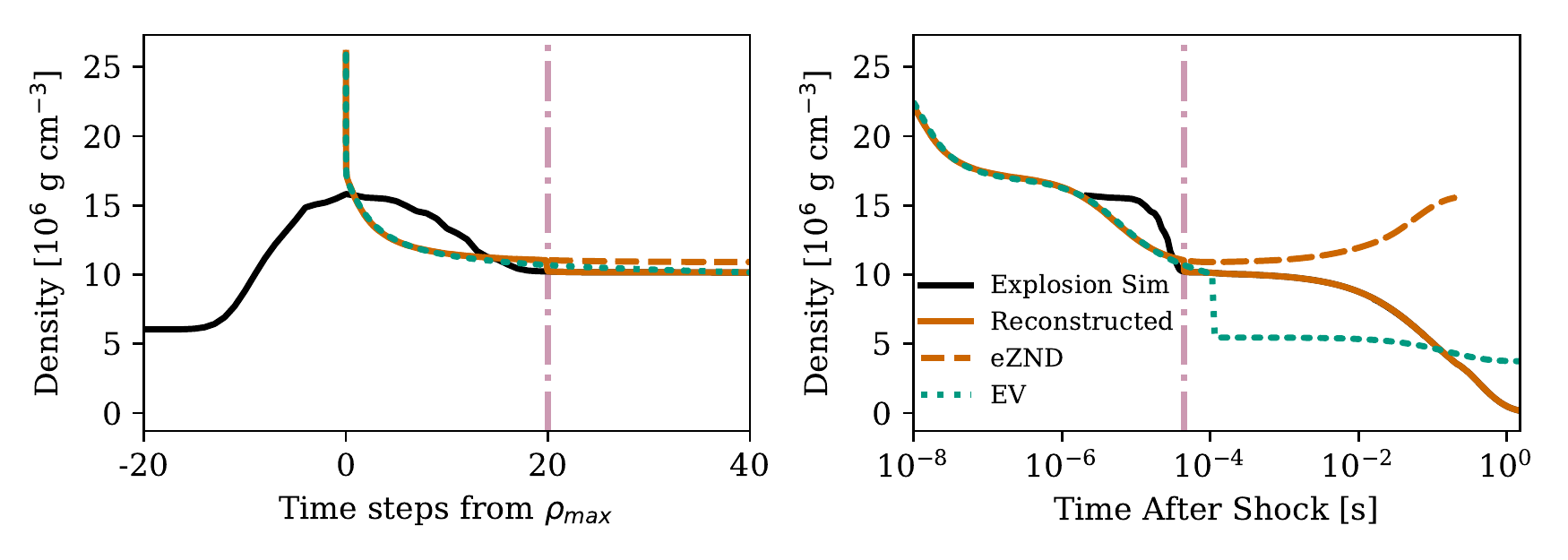}
\centering\includegraphics[scale=1]{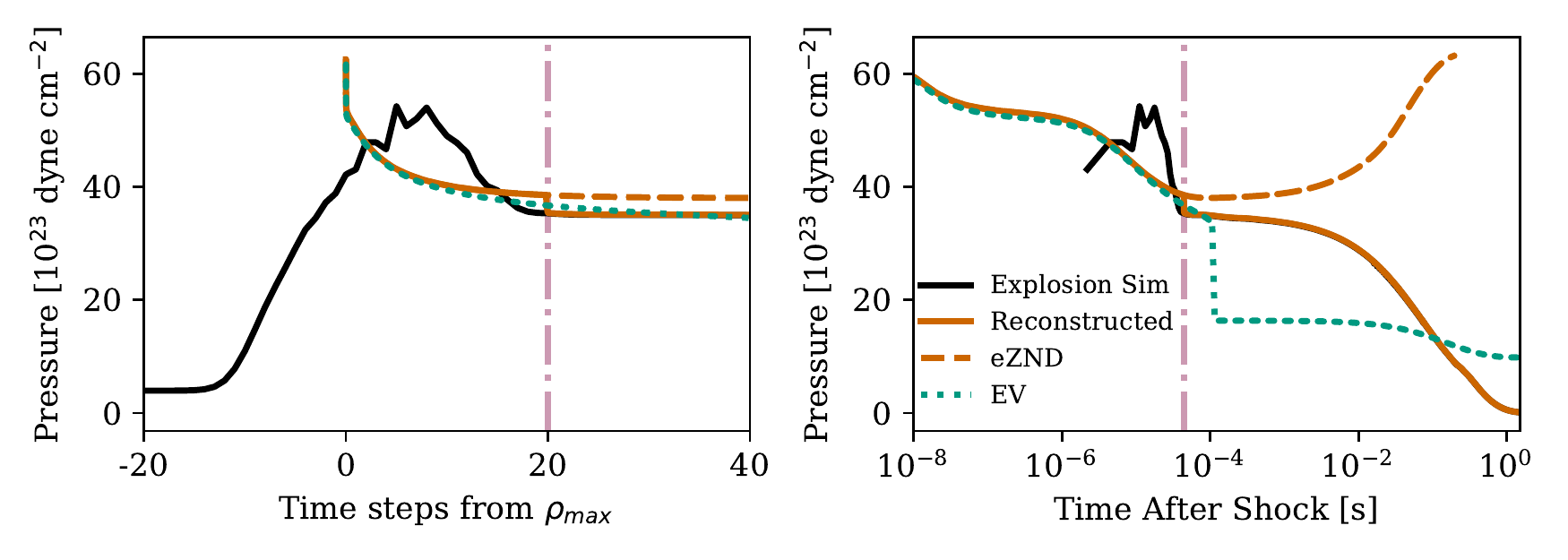}
\centering\includegraphics[scale=1]{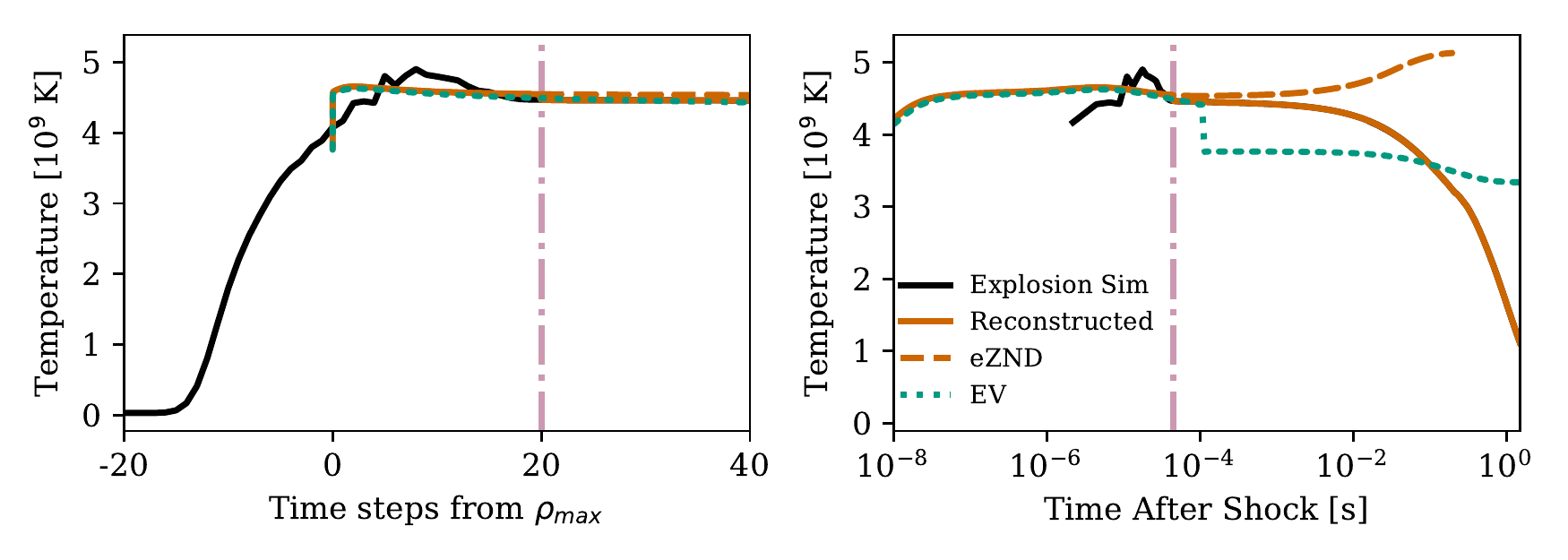}
\centering\caption{Example of the reconstructed density (top), pressure (middle), and temperature (bottom) histories of a track with a pre-detonation density of 6$\times$10$^{6}$~g~cm$^{-3}$ from the 0.0625 km resolution calculation. The black line represents the values directly from the explosion simulation, the orange line represents the full reconstructed history, the orange, dashed line represents the continuation beyond the pasting point of the structure of a fully steady-state overdriven detonation, and the green line shows the results of an eigenvalue detonation at this density. Left shows the histories as function of timestep prior to and after the density peak and right shows the histories as a function of time after shock on a log scale. }
\label{fig:temp_recon}
\end{figure*}

Figure~\ref{fig:temp_recon} shows an example of the reconstruction of a track with a pre-detonation density of 6$\times$10$^{6}$~g~cm$^{-3}$ from the 0.0625 km resolution explosion simulation.
The black lines show the recorded histories from the explosion simulation, the orange, solid lines show the reconstructed history of the particle, the orange, dashed lines show the continuation beyond the pasting point of the structure of a fully steady-state overdriven detonation, and the green dotted line shows the results of an eigenvalue detonation at this density.
The pasting point is shown as the vertical dash-dotted line and is 20 time steps after the density peak.
The unresolved structure of the detonation can seen clearly in the middle panel of figure~\ref{fig:temp_recon}.
What should be a sharp, instantaneous peak in pressure is a broad, multi-peaked feature created by the broad, unresolved detonation front.
The pressure peak from the eZND calculation is sharper, and reaches a higher maximum value than the pressure peak seen in the explosion simulation.
The unresolved detonation imprints similar features in both the density (top) and temperature (bottom).
The reconstructed detonation is slightly stronger than the eigenvalue detonation, and at the pasting point the eigenvalue detonation predicts values of pressure, density, and temperature that more closely match the values seen in the explosion simulation.

The steady-state portion of our quasi-steady-state reconstruction using the simulation detonation speed over-predicts the density and pressure by around 10$\%$ and temperature by 1$\%$ at the pasting point.
We consider this a satisfactory match, given the degree to which the simulation is under-resolved, and will treat this mismatch as a component of the remaining uncertainty in our final yields.
Reducing this uncertainty is left to future work.
Improved modeling of the unresolved processes in the simulation will likely be necessary, and a better approximate solution of the quasi-steady-state structure in the presence of a density gradient may also be required. 


\section{Comparison of Nucleosynthetic Results}
\label{sec:results}

\begin{figure*}
	\centering\includegraphics[scale=.90]{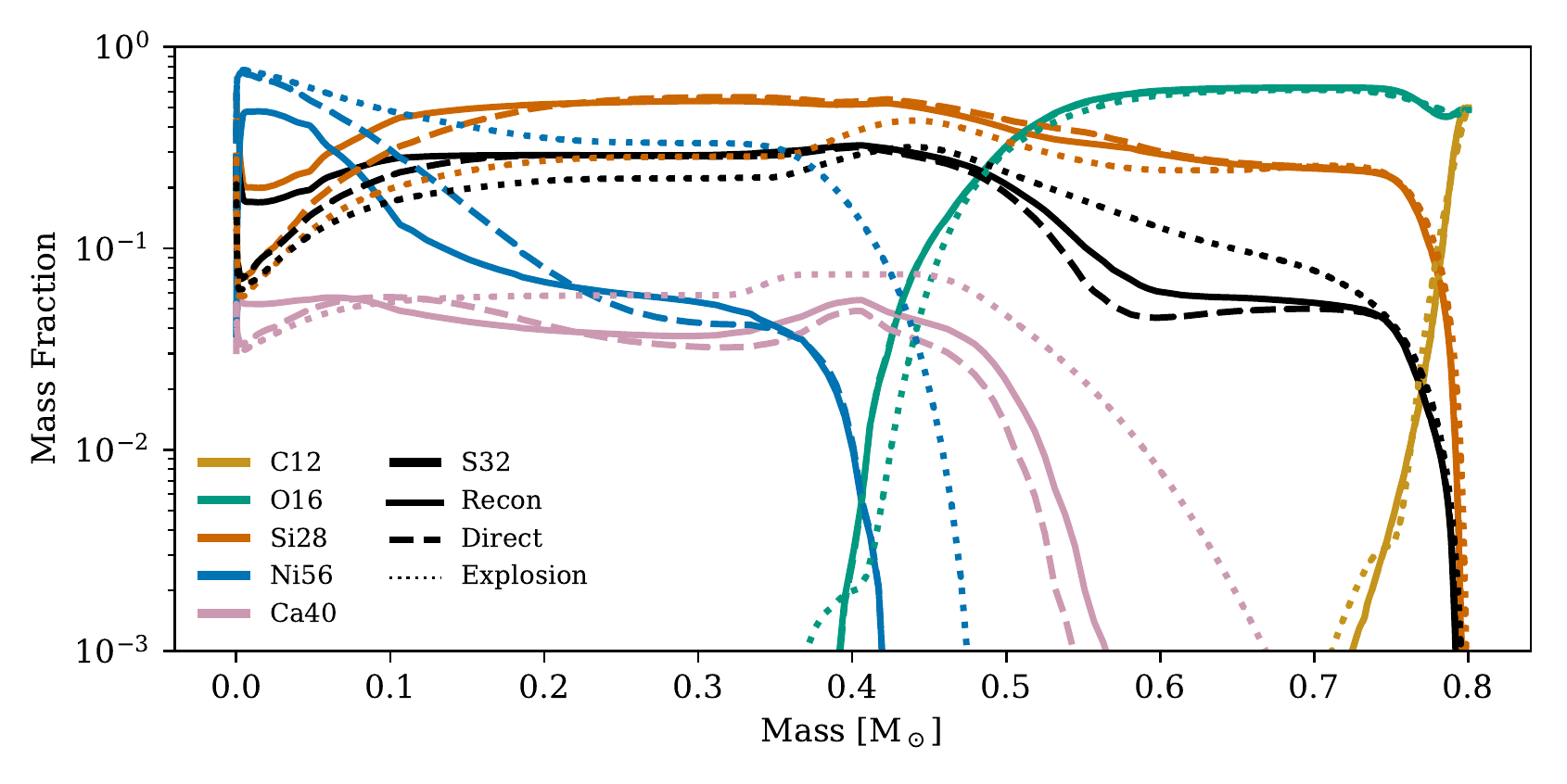}
	\centering\includegraphics[scale=.90]{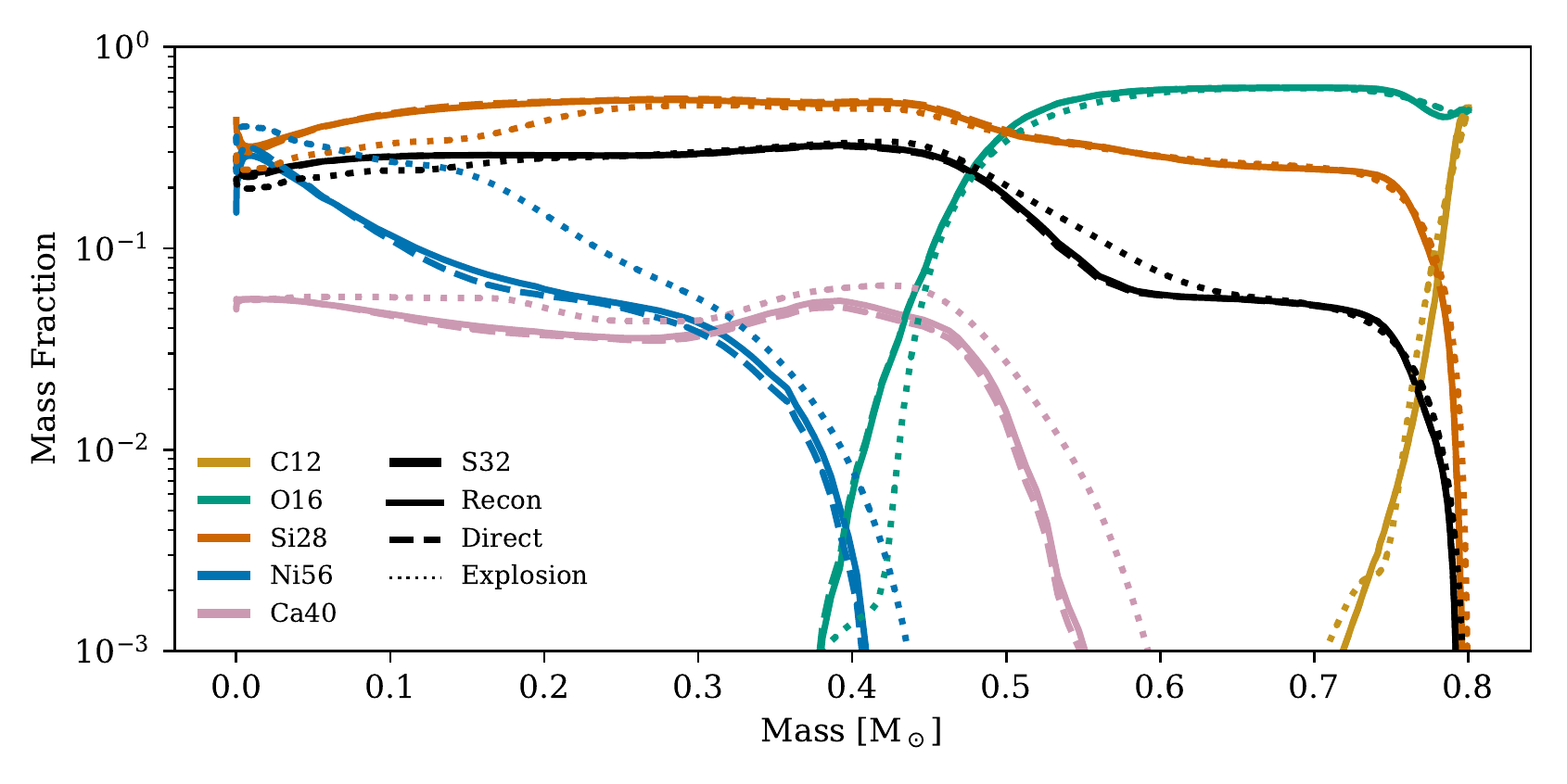}
	\caption{\label{compare_profile}
		Isotopic mass fractions vs integrated mass coordinate for two resolutions, 2~km (top), 0.0625 km (bottom). The results from the explosion simulation are represented by the dotted, the results from direct post-processing by the dashed line, and the results from detonation reconstruction by the solid line.}
\end{figure*}

Given  that the detonation is unresolved through the majority of the star at even at our highest resolution, what effect will this have on the yields compared to those produced in our two post-processing methods?
Figure \ref{compare_profile} shows the mass fraction-mass profiles produced by the explosion simulations (dotted), direct post-processing (dashed), and post-processing with detonation reconstruction (solid).
The most striking difference between the produced mass fractions is the amount of $^{56}$Ni produced in the explosion simulation compared to either of the two post-processing methods.
For both resolutions, the mass fraction of $^{56}$Ni is consistently higher throughout the ejected material.
This is also seen in the mass fraction of $^{32}$S and $^{40}$Ca, but not in the mass fraction of $^{28}$Si, in which post-processing methods produce higher amounts than the explosion simulation.
The higher mass fraction of these heavier species compared to the lower mass fraction of $^{28}$Si point to burning being more complete in explosion simulation than in the post-processing.
The disparity between the explosion simulation and post-processing results shrinks with increasing resolution, meaning the burning becomes less complete as the resolution increases.
This echoes what was shown in section \ref{hydro_results}: the strength of the detonation decreased with increasing resolution.

\begin{figure*}
	\centering\includegraphics[scale=.90]{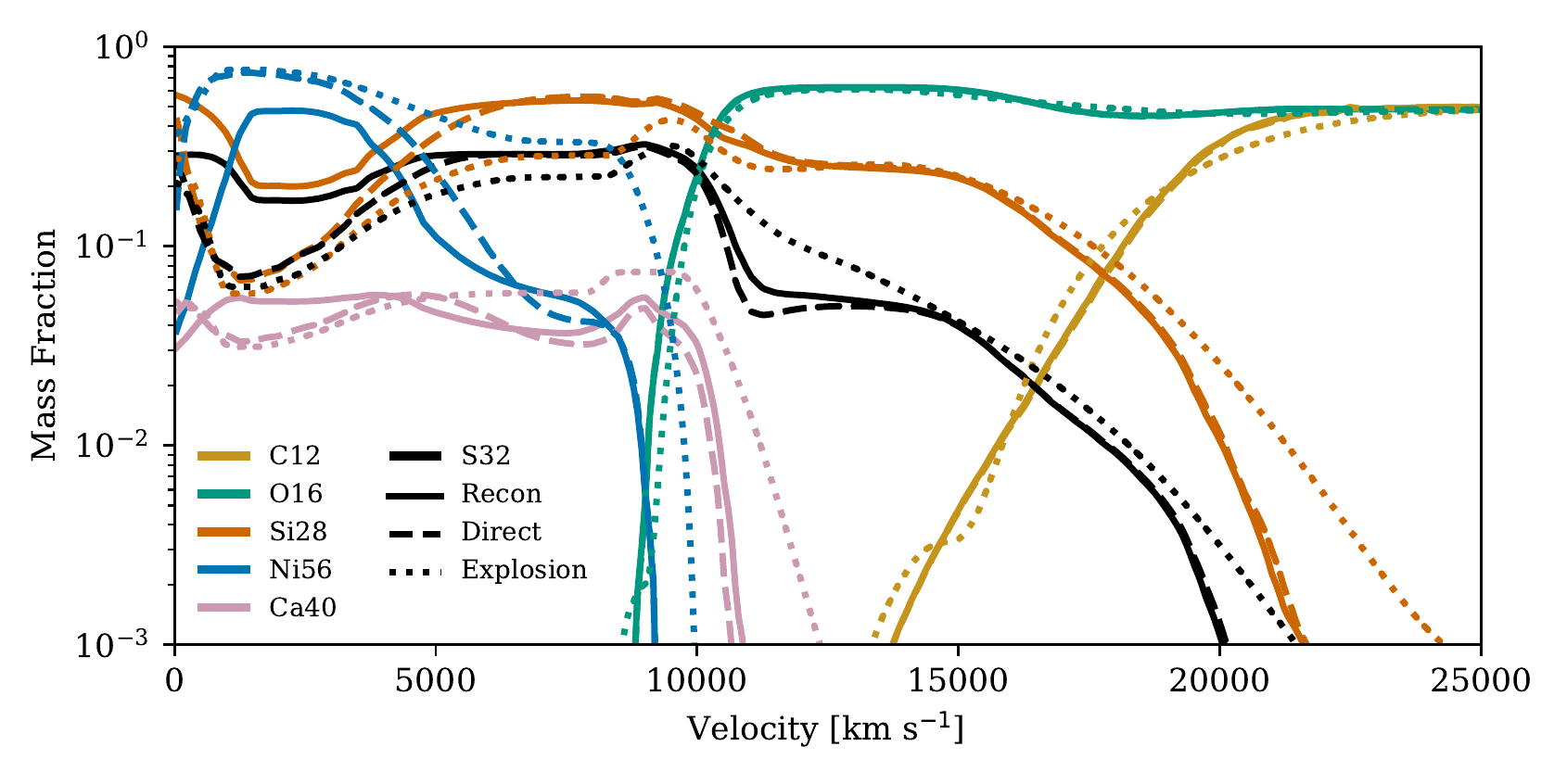}
	\centering\includegraphics[scale=.90]{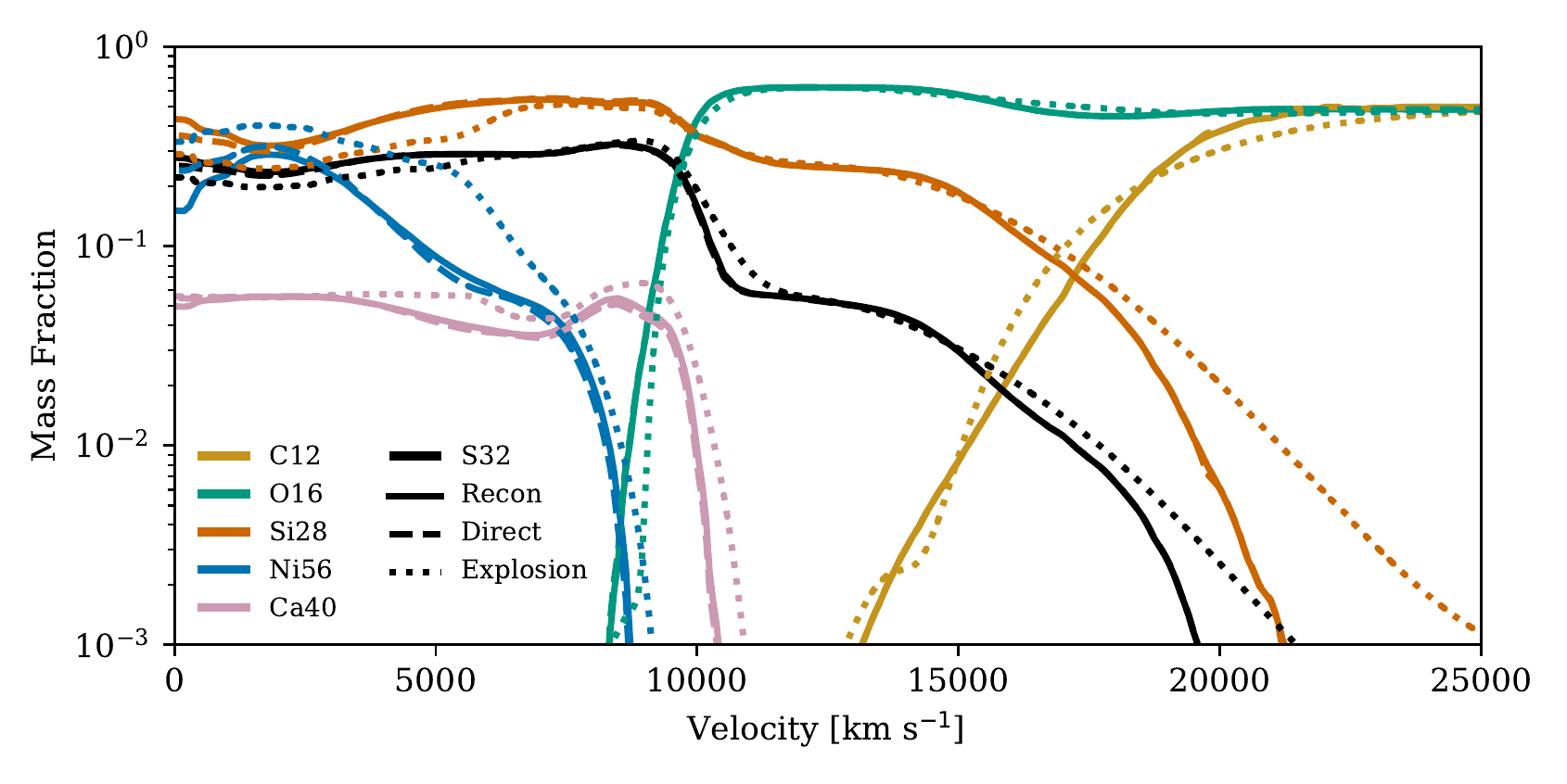}
	\caption{\label{fig:vel_profile}
		Isotopic mass fractions vs velocity for the two resolutions, 2~km (top), 0.0625~km (bottom). The results from the explosion simulation are represented by the dotted line, the results from the direct post-processing by the dashed line, and the results from the detonation reconstruction by the solid line.}
\end{figure*}

Similar features can be seen in the velocity profile (figure~\ref{fig:vel_profile}).
Much like in the mass profile, the reconstructed results lie between the explosion simulation and direct post-processing results.
However, in the velocity profile, the differences in the outer regions of the star, velocities greater than 20000 km s$^{-1}$, are much more pronounced.
The material at these velocities has experienced more complete burning than is seen in either post-processing method.
This is the region of the star where the density gradient is strongest, and the detonation is being strengthened.
The method of reconstructing the unresolved portion of the detonation with the structure of a steady state detonation may not work as well in this region.
What is likely happening is that the refinement that we force when the detonation leaves the star has a smearing effect on the isotopic mass fractions of the high velocity ejecta.
However, as this does not affect the amount of material present in this region, and the mass coordinate of the isotopic mass fractions remains constant, the total yields produced by explosion are not affected by the derefinement.

\begin{figure*}\centering{
		\includegraphics[]{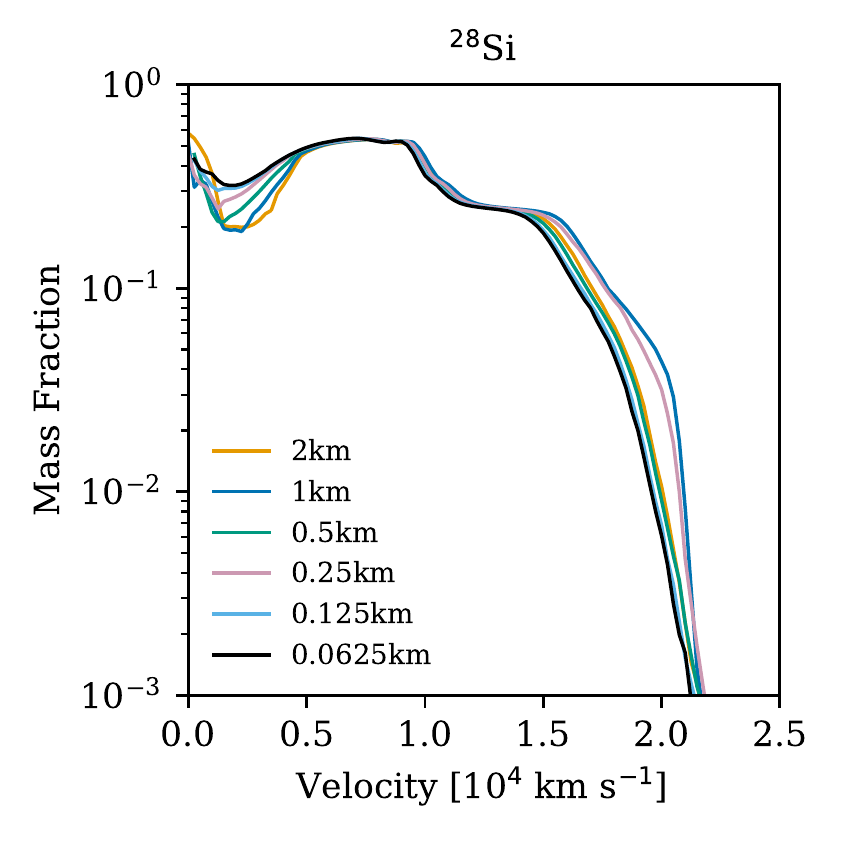}\
		\includegraphics[]{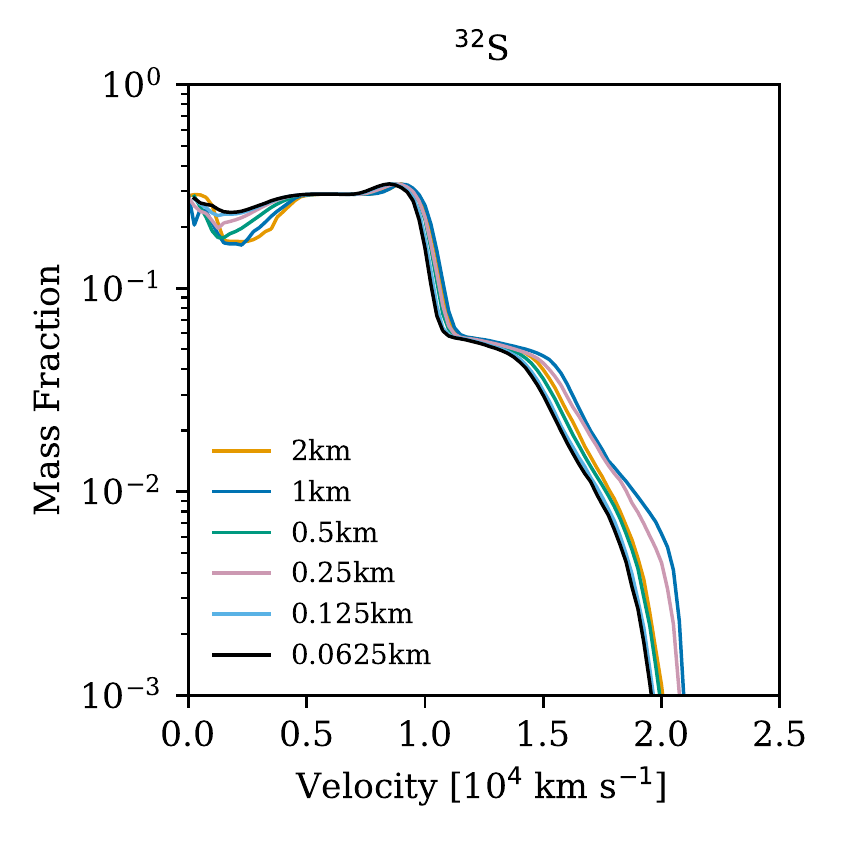}
		\includegraphics[]{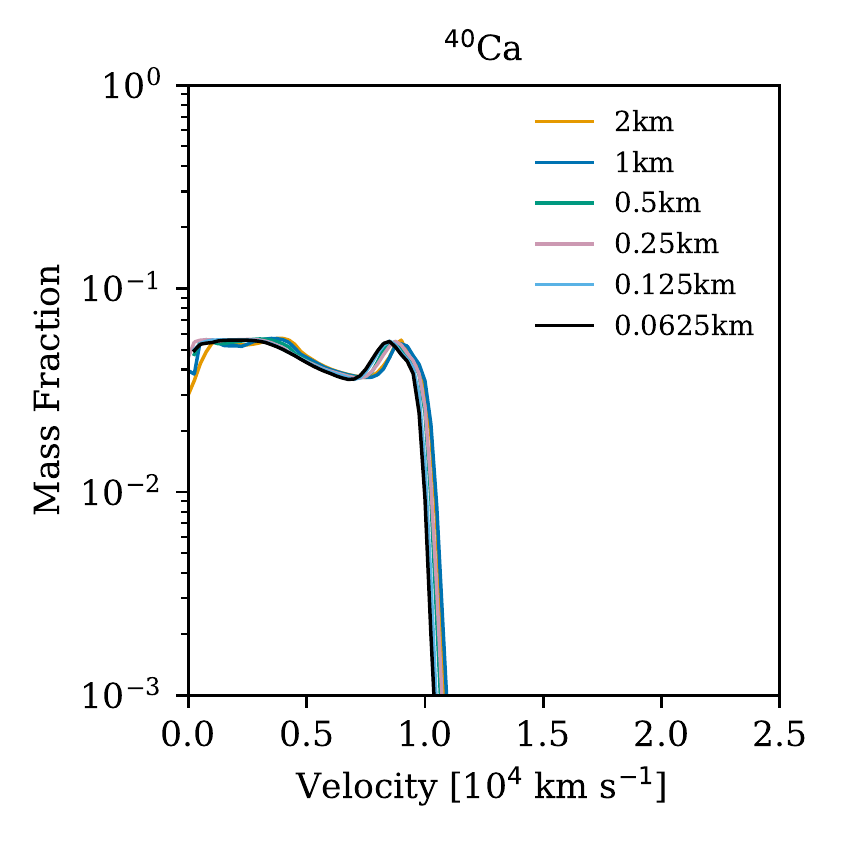}\	
		\includegraphics[]{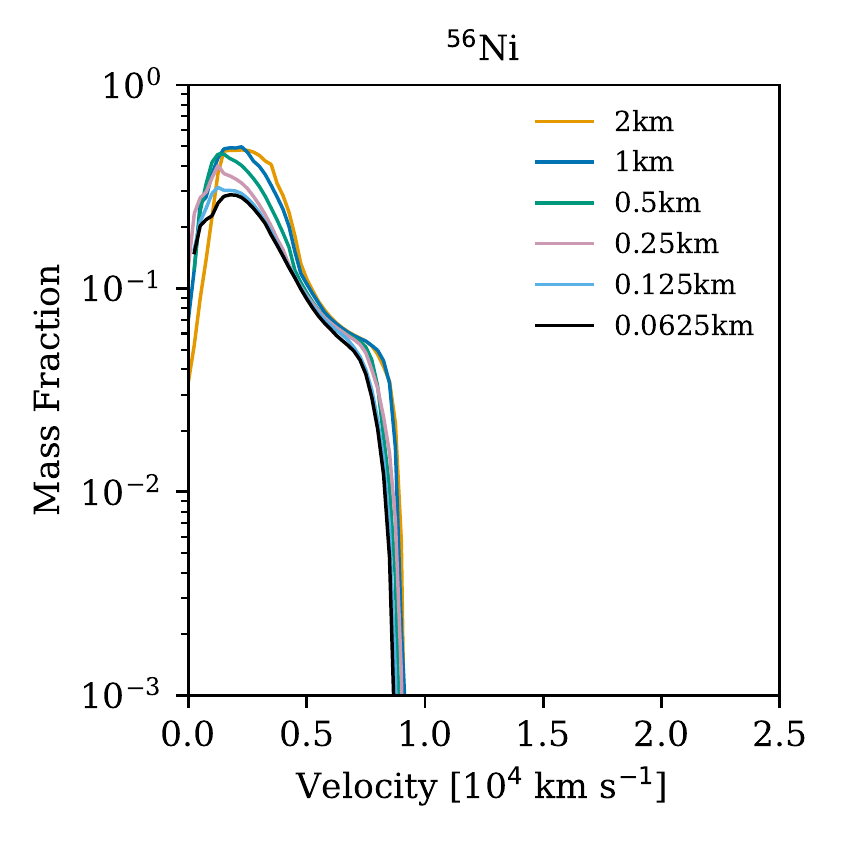}}
	\caption{\label{recon_vel_profile}
		Isotopic mass fractions of $^{28}$Si (top,left), $^{32}$S (top, right), $^{40}$Ca (bottom,left), and $^{56}$Ni (bottom,right) vs velocity from the results of post-processing with detonation reconstruction at 2 (orange), 1 (blue), 0.5 (green), 0.25 (magenta), 0.125 (cyan), and 0.0625 (black) km resolutions. 
		The greatest variations with resolution are seen in the lowest and highest velocity regions in $^{28}$Si and $^{32}$S, and at the low velocity regions in $^{56}$Ni.
	}
\end{figure*}

Figure~\ref{recon_vel_profile} shows the mass fraction of $^{28}$Si, $^{32}$S, $^{40}$Ca, and $^{56}$Ni produced by post-processing with detonation reconstruction in velocity space for the 2, 1, 0.5, 0.25, 0.125, and 0.0625 km resolution runs.
The greatest variations with resolution are seen in the mass fractions of $^{28}$Si, $^{32}$S, and $^{56}$Ni at the lowest velocities with the 0.0625 km resolution results showing the lowest average amount of $^{56}$Ni in this region.
This is unsurprising as this material is produced in the highest density region of the progenitor where the detonation speed was most sensitive to resolution.
The 1 and 0.25 km cases show discrepancies in the amount of $^{28}$Si and $^{32}$S at velocities higher than 15000~km~s$^{-1}$.
This is likely a combination of poor particle sampling in the low density regions of the progenitor and the possibility of particles being displaced into higher velocity ejecta when derefinement occurs after the shock has left the star.

Tables~\ref{tab:direct_yields}~and~\ref{tab:recon_yields} contain the integrated yields of $^{12}$C, $^{16}$O, $^{28}$Si, $^{32}$S, $^{40}$Ca, and $^{56}$Ni produced by direct post-processing and post-processing with detonation reconstruction as a function of resolution.
Across all resolutions, both methods arrive at similar values for the amount of produced $^{28}$Si.
Moving down the table to heavier isotopes, the two methods begins to diverge at lower resolutions.
At 2 km resolution, the $^{56}$Ni yields differ by $\sim$30$\%$, while at 0.0625 km the difference has fallen to $\sim$2$\%$.
Also the variation in $^{56}$Ni yield across resolutions, and therefore, the error at lower resolutions, is smaller for the reconstruction method.

\begin{figure*}
	\centering\includegraphics[]{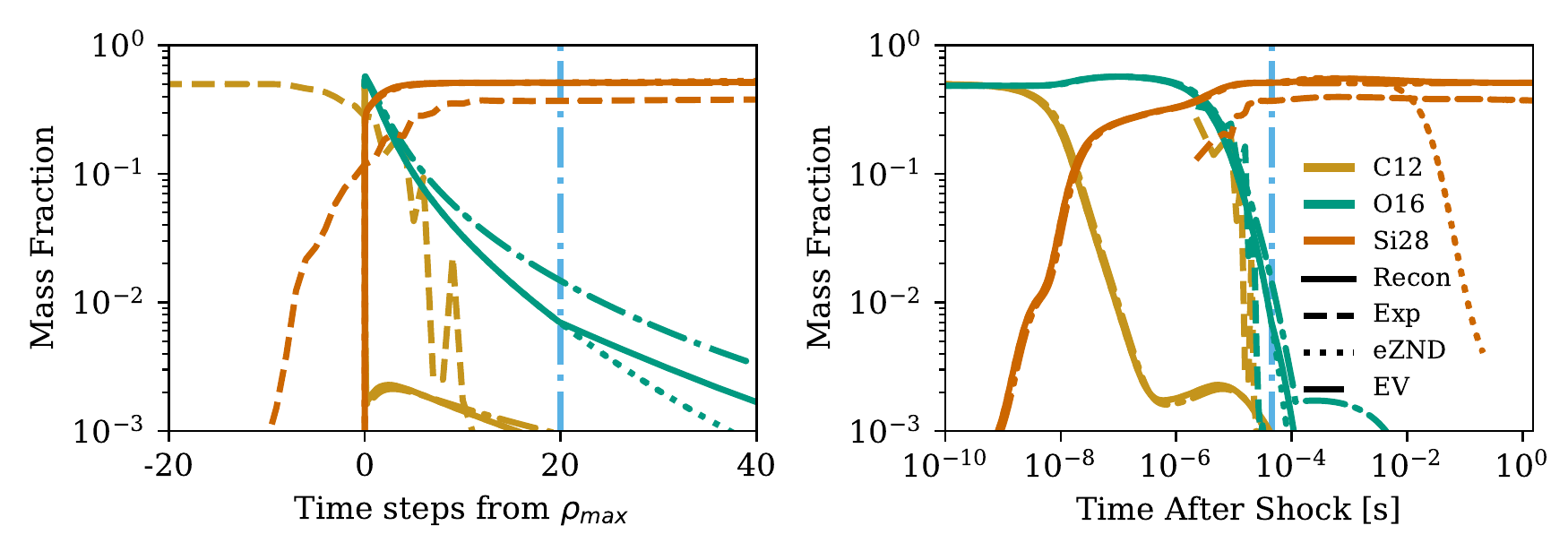}
	\caption{Mass fraction histories of $^{12}$C (gold), $^{16}$O (green), and $^{28}$Si (orange) of a track with a pre-detonation density of 6.0$\times$10$^{6}$ g cm$^{-3}$. The solid line represents the mass fractions produced by post-processing with detonation reconstruction, the dashed lines are the mass fractions taken directly from the explosion simulation, the dotted line are the mass fractions from the continuation of the steady-state detonation calculation if continued beyond the pasting point, and the dash-dotted line are the mass fractions produced by the eigenvalue detonation at this density. The vertical dash-dotted line shows the location of the pasting point. The unresolved nature of the burning in the explosion simulation can be seen in the simultaneous consumption of carbon and oxygen.}
	\label{fig:abund_example}
\end{figure*}

The unresolved detonation is a known problem.
One possible solution is to not allow burning to occur in the material within the shock.
Our explosion makes use of the "no shock burning" flag in FLASH.
If the "no shock burning" works as intended, the burning would begin in fully unburned material.
However, due to the numerical nature of the shock, mixing of partially burned and unburned material can occur before burning begins.
Figure~\ref{fig:abund_example} shows the mass fraction of $^{12}$C (gold, dashed) and $^{16}$O (green,dashed) recorded by the previously examined tracer history with a pre-detonation density of 6$\times$10$^{6}$~g~cm$^{-3}$ from 0.0625~km resolution explosion simulation as function of timesteps prior to and after the density peak (left) and as a function of time after the density peak (right).
In the left panel, both the oxygen and carbon mass fractions are beginning to fall before the tracer encounter the density peak due to numerical mixing.
It appears that turning off burning in the shock is not enough to prevent the mixing of unburned and burned material.

The unresolved nature of the burning after the density peak is much more nefarious.
In the ZND detonation structure, the pressure and density peaks, which occur at the leading shock, are separated from the temperature peak, which occurs well into the reaction region.
This separation, which governs the dynamics of the detonation propagation, is unresolved in the simulation and replaced with a structure in which the pressure peak is broadened and shifted to be approximately coincident with the temperature peak instead (figure~\ref{fig:temp_recon}).
The consequence of this can be seen in the dashed lines of figure~\ref{fig:abund_example}.
The mass fractions recorded by the tracer show that carbon and oxygen are being consumed simultaneously.
In reality, as can be seen by the reconstructed results (solid lines) in the right panel, carbon, oxygen and silicon burning occur in three distinct stages.
Mixing of these burning stages results in more complete burning.
The tracer recorded a final $^{56}$Ni mass fraction of 0.205.
The same tracer post-processed with detonation reconstruction gives a $^{56}$Ni mass fraction of 0.073, a factor of 2.8 lower.
The panel on the right also gives an insight into the extreme time resolution necessary to capture all three stages of burning.
To separate carbon and oxygen burning, it would require timesteps on the order of 10$^{-10}$ seconds.
This is untenable for full star simulations.

This brings back the concern of what effects the explosion simulation resolution would have on the post-processing results. 
In section \ref{recon}, we showed how the resolution affected the recorded temperature, density, and pressure histories of four particles varying from resolved to unresolved, but how does it affect the results of post-processing?
In figure \ref{compare_profile}, the direct post-processing and post-processing with reconstruction results do differ by a small amount.
In the 2 km resolution case, the direct post-processing is the lower limit of the $^{32}$S, $^{40}$Ca, and $^{56}$Ni mass fractions and the upper limit of the $^{28}$Si mass fraction.
This is due to what is seen in figure~\ref{fig:temp_recon}. The shock, as experienced by the particle, is broad and not as strong as it should be, resulting in a lower peak density and pressure and thus, less burning.
However, why is it not closer to the results of the explosion simulation since the identical physical conditions are integrated by an identical nuclear reaction network?
The factors discussed in the previous two paragraphs are the cause: the initial mass fractions and the unresolved burning scales.
When a tracer is post-processed, its initial mass fractions are set to be that of unburned material, removing the problem of numerical mixing.

The results of the post-processing with detonation reconstruction tend to fall between the results of the explosion simulation and the direct post-processing.
The hope with a method such as this is that it is able to produce results that do not significantly vary with changing resolution.
Table~\ref{tab:recon_yields} shows that is true for a small selection of yields except for $^{56}$Ni.
The results of the detonation reconstruction are tied to the resolution in two important ways.
One, the detonation speed used to reconstruct the structure of the detonation is taken directly from the explosion simulation.
Figure~\ref{det_speed_fig} shows how the detonation speed varies with resolution, and while the detonation speed is relatively converged in the low density portions of the progenitor, it's less certain in the high density regions where isotopes such as $^{56}$Ni are produced.
Two, though the unresolved portion of the detonation is trimmed from each track during the reconstruction, the portion of the track that the reconstructed detonation structure is stitched onto will have some dependence on the resolution.
Though it is not completely immune to resolution, the ability to include all of and resolve the important physical processes during post-processing make it a powerful technique. 
Also, since the pasting point location and the speed used in the reconstruction are exposed parameters, they can be evaluated or varied to assess systematic uncertainty.
This is discussed more in section~\ref{sec:uncertainty}.
Evaluation of uncertainty in this way can be done without the high-resolution comparisons performed here, which are not feasible for multi-dimensional simulations.

\begin{deluxetable*}{c|cccccc}
	\tablewidth{\columnwidth}
	\tablecaption{Selected Direct Post-Processing Yields by Resolution in M$_\odot$ \label{tab:direct_yields}}
	\tablehead{%
		\colhead{Iso} & \colhead{2km} & \colhead{1km} & \colhead{0.5km} & \colhead{0.25km} & \colhead{0.125km} & \colhead{0.0625km}
	}
	\startdata
	$^{28}$Si & 0.301 & 0.314 & 0.316 & 0.326 & 0.322 & 0.322  \\
	\hline
	$^{32}$S\phm{.}\Tstrut & 0.144 & 0.153 & 0.155 & 0.161 & 0.152 & 0.160 \\
	\hline
	$^{40}$Ca\Tstrut & 0.0206 & 0.0214 & 0.0212 & 0.0215 & 0.0212 & 0.0211 \\
	\hline
	$^{56}$Ni\Tstrut & 0.0785 & 0.0642 & 0.051 & 0.0418 & 0.0365 & 0.0344
	\enddata
\end{deluxetable*}
\begin{deluxetable*}{c|ccccccc}
	\tablewidth{\columnwidth}
	\tablecaption{Selected Yields from Post-Processing with Detonation Reconstruction by Resolution in M$_\odot$ \label{tab:recon_yields}}
	\tablehead{%
		\colhead{Iso} & \colhead{2km} & \colhead{1km} & \colhead{0.5km} & \colhead{0.25km} & \colhead{0.125km} & \colhead{0.0625km}
	}
	\startdata
	$^{28}$Si & 0.308 & 0.316 & 0.315 & 0.322 & 0.319 & 0.319 \\
	\hline
	$^{32}$S\phm{.}\Tstrut & 0.158 & 0.162 & 0.161 & 0.165 & 0.161 & 0.161\\
	\hline
	$^{40}$Ca\Tstrut & 0.0226 & 0.0229 & 0.0225 & 0.0226 & 0.0221 & 0.0220\\
	\hline
	$^{56}$Ni\Tstrut & 0.0536 & 0.0502 & 0.0426 & 0.0386 & 0.0359 & 0.0350
	\enddata
\end{deluxetable*}

\section{Comparisons to Alternate Methods}
Previous studies of SNe Ia, such as \cite{miles_2016}, have made use of post-processing, with large nuclear reaction networks, the results of explosion simulations that utilized either smaller reaction networks or parameterized burning schemes.
More recently, in an attempt to solve the problem of the unresolved detonation structure, \cite{Shen_18}, following up on related methods discussed by \cite{Kushnir_2013}, used a burning limiting method to thicken the detonation front, and then post-processed the result using a large nuclear reaction network.
In this section, we will use the results of our highest resolution calculations with the 205-species reaction network, post-processed using detonation reconstruction as the benchmark to which the results of the other methods are compared.

\subsection{Comparison to Aprox13 Plus Post-Processing}
\label{sec:aprox13}

Aprox13 is a 13-isotope alpha chain reaction network consisting of $^{4}$He, $^{12}$C, $^{16}$O, $^{20}$Ne, $^{24}$Mg, $^{28}$Si, $^{32}$S, $^{36}$Ar, $^{40}$Ca, $^{44}$Ti, $^{48}$Cr, $^{52}$Fe, and $^{56}$Ni.
It should be noted that alpha chain networks are only reasonably valid in environments where Y$_e$ does not differ greatly from 0.5.
Generally, alpha-chain reaction networks only track the ($\alpha$,$\gamma$) and ($\gamma$,$\alpha$) reactions linking the members together.
However, at the temperatures present in SNe Ia, ($\alpha$,p)(p,$\gamma$) begin to dominate, and in order to accurately capture energy generation rates and yields, these reactions need to be included.
Aprox13 is differs from a strict alpha-chain network in two ways.
First, it includes the reaction rates of $^{12}$C+$^{12}$C, $^{12}$C+$^{16}$O, and $^{16}$O+$^{16}$O.
Second, it also contains 8 of the previously mentioned ($\alpha$,p)(p,$\gamma$) rates.
These are effective rates, meaning the intermediate products, such as $^{27}$Al between $^{24}$Mg and $^{28}$Si, are not explicitly included.
This is done by assuming that the flow of protons into each channel is equal the number of protons flowing out \citep{Timmes_aprox13}.

For this comparison, we run the simulation of the explosion of a 50/50 carbon-oxygen white dwarf at 0.125~km resolution.
We include 10000 Lagrangian tracer particles to record the temperature and density history of the fluid for post-processing.
Due to the limited nature of aprox13, we cannot include metallicity during the explosion simulation.
The temperature-density histories are integrated using the MESA one-zone burner using the 205-isotope network discussed in section~\ref{sec:progen}.
The initial mass fraction of each particle is set to be 50 percent C, 48.6 percent O, and 1.4 percent metals at solar mass fractions where contributions from C,N,O have been converted to $^{22}$Ne.

\begin{figure*}
\centering\includegraphics[scale=0.85]{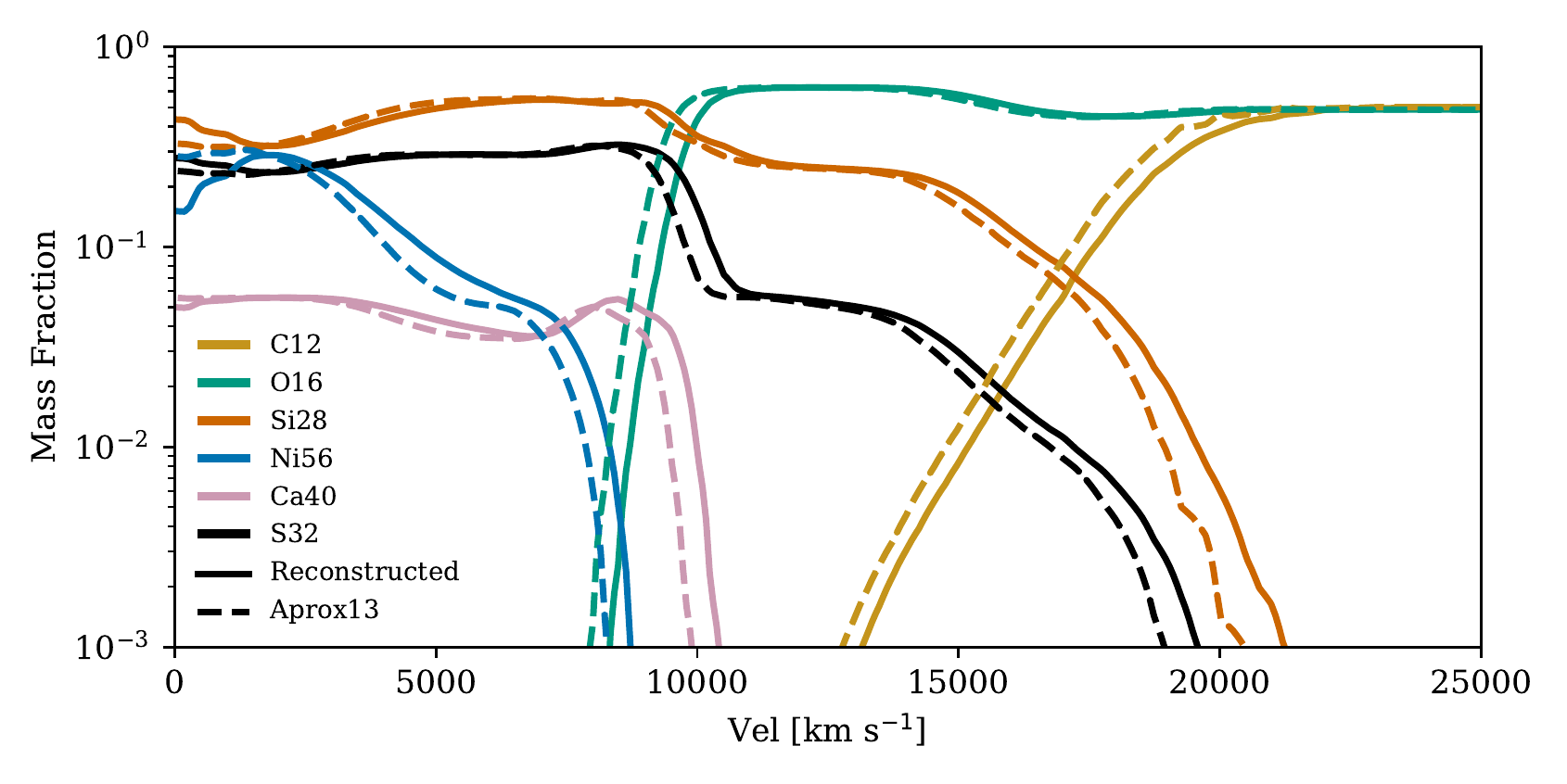}
\caption{Mass fraction vs velocity profile for the results of the post-processing with detonation reconstruction (solid) and the results of the post-processed aprox13 results (dashed)}
\label{fig:aprox13_vel_comapare}
\end{figure*}

Figure~\ref{fig:aprox13_vel_comapare} shows the post-processed results of the aprox13 calculation compared to our benchmark yields.
From 0~km~s$^{-1}$ to 2500~km~s$^{-1}$, the post-processed aprox13 results produce more $^{56}$Ni than the detonation reconstruction results.
At 2500~km~s$^{-1}$, the yields produced by the two methods are similar for the $^{28}$Si, $^{32}$S, $^{40}$Ca, and $^{56}$Ni. 
From 2500~km~s$^{-1}$ to 8000~km~s , the mass fraction of $^{56}$Ni is higher for the detonation reconstruction case, the mass fraction of of $^{28}$Si and $^{32}$S from the two methods are similar, and the post-processed aprox13 results produce slightly more $^{40}$Ca.
The Ca peak at 8000~km~s$^{-1}$ is both higher and wider in the case of the detonation reconstruction results.
The transition from the production of intermediate mass elements to primarily O and Si happens at a lower velocity for the aprox13 results.
This is also true of the transition to carbon burning and finally, extinction.

\begin{figure*}
\centering\includegraphics[scale=0.85]{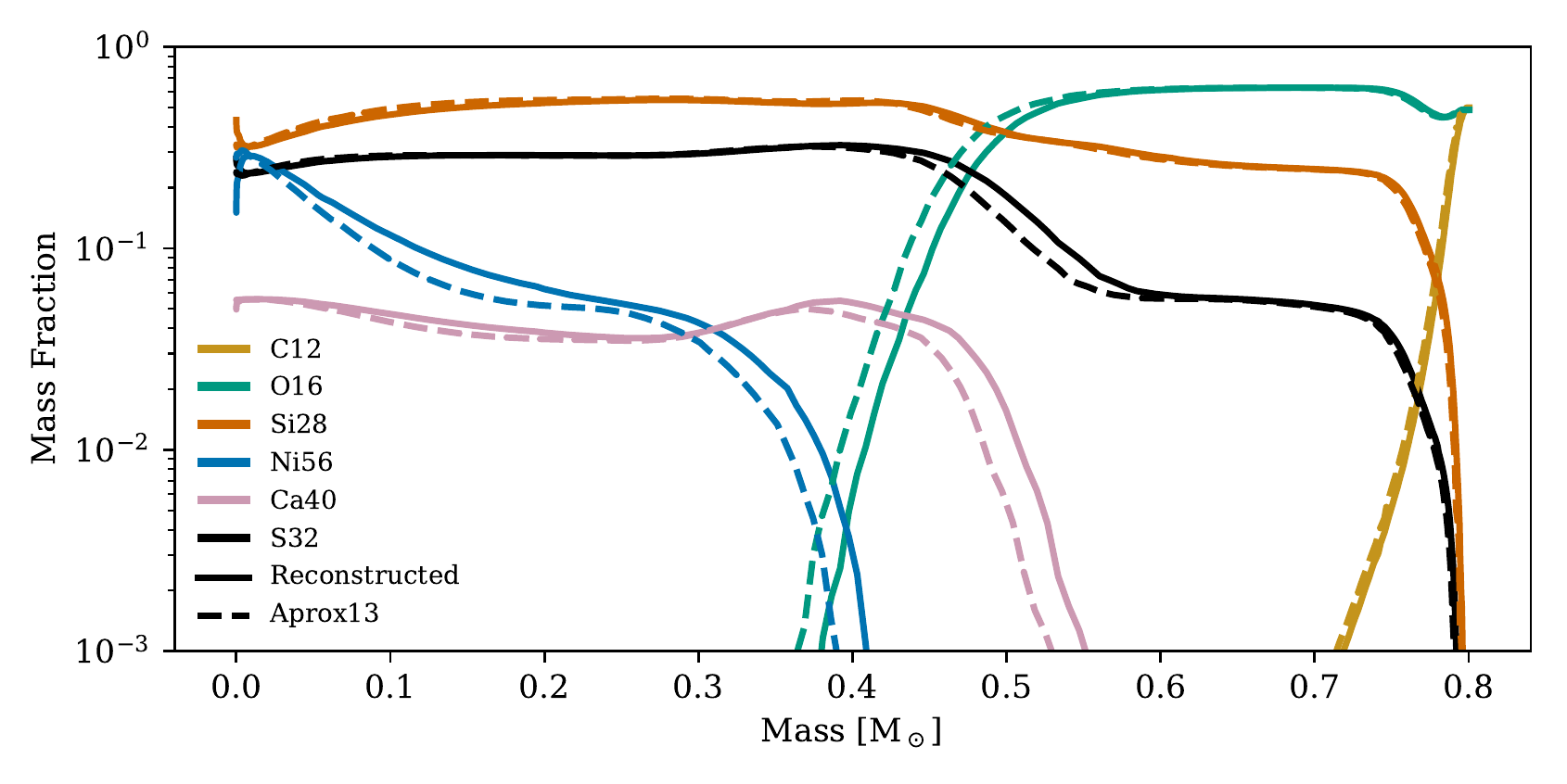}
\caption{Mass fraction vs mass profile for the results of the post-processing with detonation reconstruction (solid) and the results of the post-processed aprox13 results (dashed)}
\label{fig:aprox13_mass_comapare}
\end{figure*}

Figure~\ref{fig:aprox13_mass_comapare} shows the post-processed results of the aprox13 calculation compared to the results of the post-process with detonation reconstruction in integrated mass.
In mass, the results are not dissimilar from the velocity profile.
The reconstruction method produces more completely burned material to further extents than the post-processed aprox13 results.
However, from about 0.6 M$_\odot$ to edge of the star, the results of the two methods are similar.

Table~\ref{tab:aprox13_yields} gives the yields from both methods.
The detonation reconstruction method produces more $^{32}$S, $^{40}$Ca, and $^{56}$Ni than the post-processed aprox13 results.
This is in-line with what is seen in the integrated mass profiles, where the results of the detonation reconstruction method are more completely burned through more of the star.

\begin{deluxetable}{c|cc}
	\tablewidth{\columnwidth}
	\tablecaption{Yield in M$_\odot$ Comparison: Post-processing with Detonation Reconstruction vs aprox13 + Direct Post-processing \label{tab:aprox13_yields}}
	\tablehead{%
		\colhead{Iso} & \colhead{Reconstruction} & \colhead{Aprox13 + Post-Processing}
	}
	\startdata
	$^{28}$Si & 0.319 & 0.323 \\
	\hline
	$^{32}$S\phm{.}\Tstrut & 0.161 & 0.156 \\
	\hline
	$^{40}$Ca\Tstrut & 0.0220 & 0.0199 \\
	\hline
	$^{56}$Ni\Tstrut & 0.0350 & 0.0295
	\enddata
\end{deluxetable}

\subsection{Parameterized Burning with Post-Processing}
Parameterized burning schemes are a computationally inexpensive method of capturing the energetics and propagation of burning processes in SNe Ia, but due to their inability to track the mass fractions of individual species, post-processing is required for the calculation of nucleosynthetic yields.
Here we make comparisons to the parameterized burning scheme of \cite{townsley_2016}.
In the Townsley16 parameterized burning scheme, nuclear burning is tracked through three progress variables measuring the completion of three consecutive processes.
The first corresponds roughly to carbon consumption, the initial conversion of fuel to ash, $\phi_{fa}$.
The second, $\phi_{aq}$, represents the conversion of this ash to quasi-(statistical)-equilibrium (QSE) silicon-dominated material via oxygen consumption.
Finally, $\phi_{qn}$, represents full relaxation from QSE to nuclear statistical equilibrium (NSE) by the consumption of silicon group material to produce iron group material.

The endpoint energy released is calculated based on a table of the NSE state \citep{Caldetal07,townsley.calder.ea:flame,SeitTownetal09}.
The timescales for oxygen and silicon consumption are temperature-dependent fits to the results of reaction network or detonation computations.
These were originally based on those of \citet{Khokhlov1991Delayed-detonat}, but re-computed by \citet{Caldetal07}.
However, \citet{townsley_2016} found the treatment of silicon consumption inadequate to reproduce the structure of planar detonations, and so re-calibrated the model based on planar detonations computed using the ZND formalism.
The oxygen consumption stage was left unchanged in Townsley16.

For this comparison, we calculate the explosion of a 0.8 M$_\odot$ progenitor with initial mass fractions of 50$\%$ $^{12}$C, 48.6$\%$ $^{16}$O, and 1.4$\%$ $^{22}$Ne using the Townsley16 parameterized burning scheme.
$^{22}$Ne is the stand-in for the neutron excess from the metallicity in this scheme.
The explosion simulation was run at 0.125 km resolution for a total time of 7.5 seconds.
As before, 10000 tracer particles were included for post-processing purposes.
The temperature-density histories from each track were post-processed using the MESA one zone burner with the same 205-isotope nuclear reaction network used in the previous sections' calculations.
Initial mass fractions of each track were set to 50$\%$ $^{12}$C, 48.6$\%$ $^{16}$O, and 1.4$\%$ metals with mass fractions set to solar values with contributions from C,N, and O converted to $^{22}$Ne.

\begin{figure*}
	\centering\includegraphics[scale=0.85]{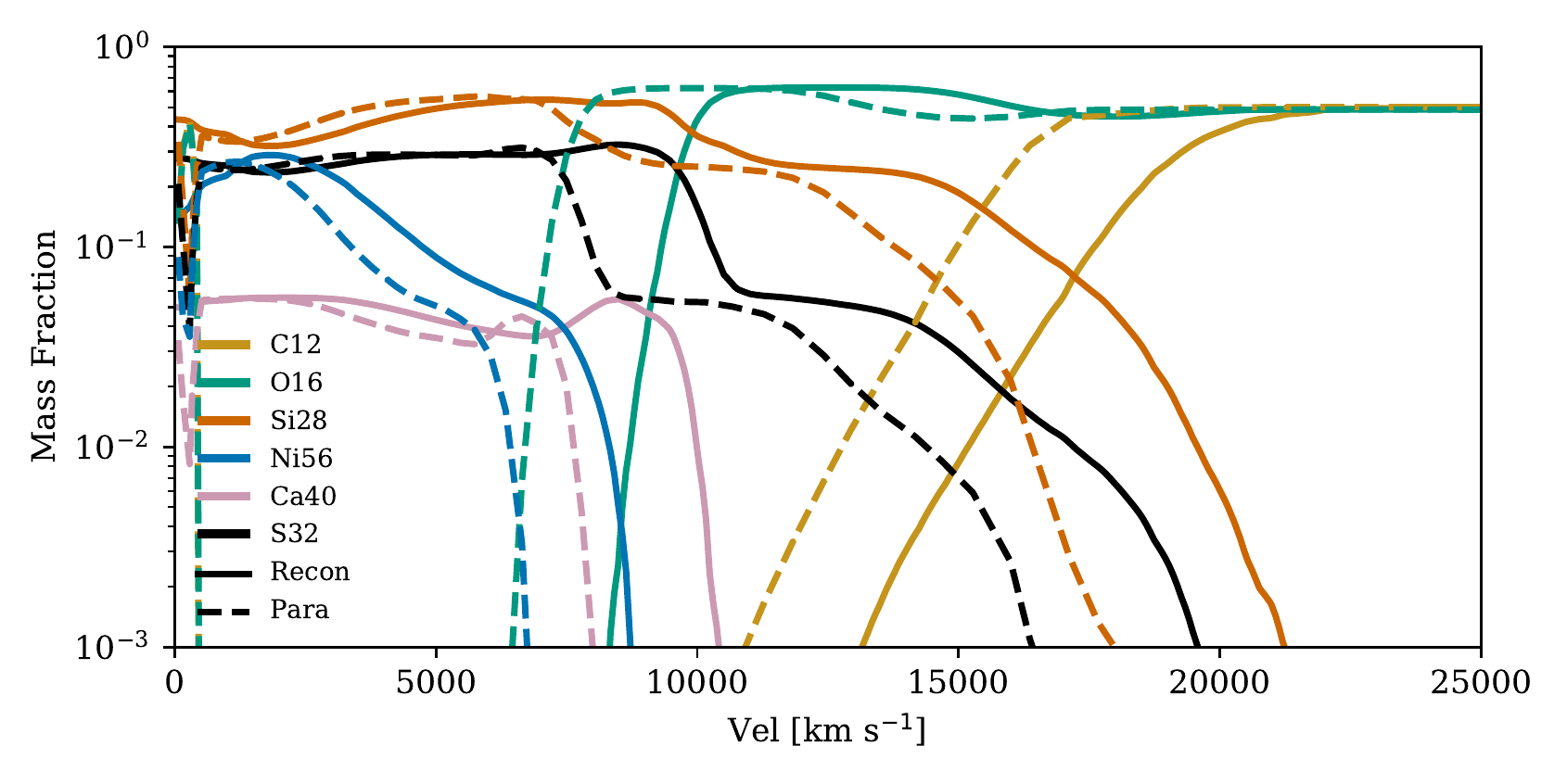}
	\caption{Mass fraction vs velocity profile for the results of the post-processing with detonation reconstruction (solid) and the results of the post-processed parameterized burning results (dashed)}
	\label{fig:para_vel_comapare}
\end{figure*}

Figure~\ref{fig:para_vel_comapare} shows the results of the parameterized burning plus post-processing (dashed lines) compared to the results of the full network plus post-processing with detonation reconstruction (solid lines) in velocity space.
The peak mass fraction of both $^{40}$Ca and $^{56}$Ni are lower in the case of the parameterized burning plus post-processing compared to the results with detonation reconstructions, while the peak mass fractions of $^{16}$O, $^{28}$Si, and $^{32}$S reach similar values.
The shape of the mass fraction profiles for each isotope produced by the two methods is similar.
However, the profiles produced by the detonation reconstruction method extend out to higher velocities.
Similarly, compared to the post-processed aprox13 results, the detonation reconstruction method predicts more complete burning at higher velocities than the post-processed parameterized burning results.
The production of $^{40}$Ca and $^{56}$Ni extends $\sim$2000 km s$^{-1}$ higher, and the production of $^{28}$Si and $^{32}$S extends $\sim$4000 km s$^{-1}$ higher. 
These large differences in velocity space could show up as large differences in spectra that could be produced from these two methods.

\begin{figure*}
	\centering\includegraphics[scale=0.85]{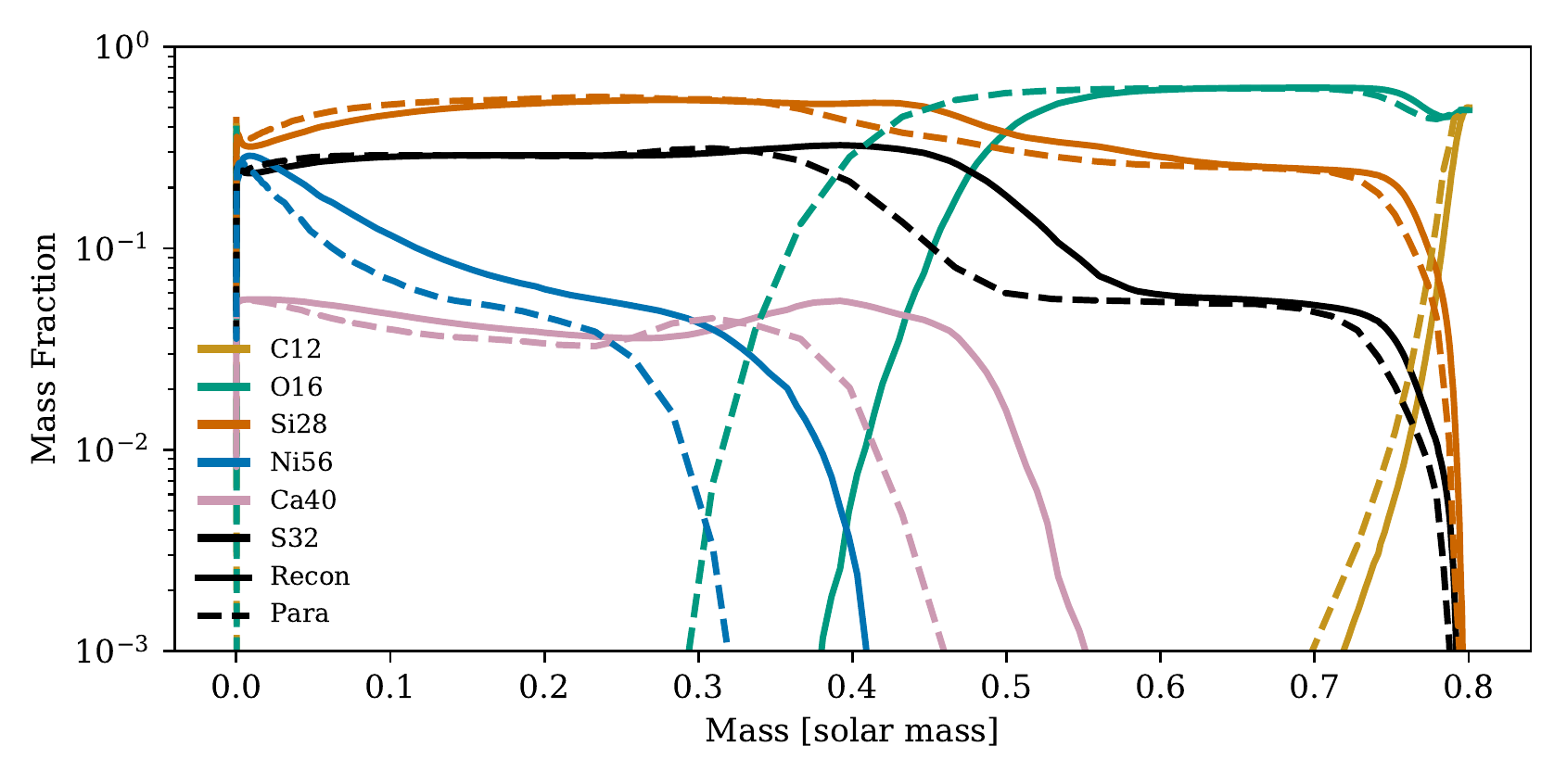}
	\caption{Mass fraction vs mass profile for the results of the post-processing with detonation reconstruction (solid) and the results of the post-processed parameterized burning results (dashed)}
	\label{fig:para_mass_comapare}
\end{figure*}

Similar analysis can be drawn from figure~\ref{fig:para_mass_comapare}.
Burning is more complete at a higher mass coordinate for the detonation reconstruction results, with $^{40}$Ca and $^{56}$Ni production occurring $\sim$0.1 M$_\odot$ further out for the detonation reconstruction method. 
Though $^{28}$Si and $^{32}$S were produced out to much higher velocities, those high-velocity mass fractions will not have a large effect on the yields.
The mass fraction of $^{32}$S is quite different from $\sim$0.35 M$_\odot$ to 0.6 M$_\odot$ between the two methods, with the detonation reconstruction method producing higher mass fraction.
The differences in the $^{28}$Si profiles are less pronounced.
The detonation reconstruction method produces slightly more in that same mass range, but at a lower magnitude.
The overall effects these differences have on the total yields can be seen in table~\ref{tab:para_yields}.
The $^{28}$Si yield differs by 5$\%$, $^{32}$S by 19$\%$, $^{40}$Ca by 41$\%$, and $^{56}$Ni by 64$\%$. 

Given the difficulty with the silicon consumption stage in planar detonations, \citet{townsley_2016} only considered $^{56}$Ni yields verified to somewhat better than 10\% for explosions that produce a significant amount of fully processed material.
This 0.8 M$_\odot$ explosion produces essentially no fully processed material, having no regions in which the mass fraction of $^{56}$Ni is larger than that of $^{28}$Si.
Thus it is somewhat expected that the parameterized model will perform poorly in this low-density limit.
The comparison here shows that, similarly to the silicon-consumption, the oxygen- and carbon-consumption stages would need to be re-calibrated for locally accurate yields at the lower densities.

The current work grew partially out of the need for verification and calibration of the parameterized burning model for curved detonations at lower densities.
As a result, the 0.8~M$_\odot$ case is, by design, particularly revealing of these deficiencies.
However, given the importance of curvature and shock-strengthening due to density gradients, the effects of which vary in different locations within the star, it is unclear if a global temperature-dependent fit to burning timescales, as used in this parameterization so far, will be sufficient to reach verifiable uncertainties better than 1\%.
Other local detonation parameters may need to be included in the model.
 
\begin{deluxetable}{c|cc}
	\tablewidth{\columnwidth}
	\tablecaption{Yield in M$_\odot$ Comparison: Post-processing with Detonation Reconstruction vs Parameterized Burning + Direct Post-processing \label{tab:para_yields}}
	\tablehead{%
		\colhead{Iso} & \colhead{Reconstruction} & \colhead{Para + Post-Processing}
	}
	\startdata
	$^{28}$Si & 0.319 & 0.303 \\
	\hline
	$^{32}$S\phm{.}\Tstrut & 0.161 & 0.135 \\
	\hline
	$^{40}$Ca\Tstrut & 0.0220 & 0.0155 \\
	\hline
	$^{56}$Ni\Tstrut & 0.0350 & 0.0213
	\enddata
\end{deluxetable}

\subsection{Shen et al 2018 Burning Limiter with Post-Processing}
\label{sec:shen}
In contrast to the detonation reconstruction method presented in this work, where we start with an unresolved calculation and then reconstruct the important pieces of physics in time and space during post-processing, \cite{Shen_18}, attempt to spatially resolve these processes during the explosion simulation via artificially thickening the detonation front.
This is accomplished by taking advantage of the split-operator nature of the FLASH code.
FLASH can solve the hydrodynamics equations and nuclear reaction equations separately using two different timesteps, $\Delta$t$_{hydro}$ and $\Delta$t$_{burn}$.
Shen18 thickens the detonation front by putting a limit on the value of $|\Delta$ln $T|_{max}$, the relative amount the temperature can change within each cell during one time step due to nuclear burning.
This is related to $\Delta$t$_{burn}$ by $\Delta$ln $T \sim \bar{\epsilon}\Delta$t$_{burn}/c_VT$, where $\bar{\epsilon}$ is the average energy generation rate over $\Delta$t$_{burn}$ and $c_V$ is the specific heat at constant volume.
During a FLASH timestep, if $\Delta$ln $T > |\Delta$ln $T|_{max}$ then a new value of $\Delta$t$_{burn}$ is found via a root find, but $\Delta$t$_{hydro}$ is left unchanged.

We used 150 tracer particles produced from the 0.5 km resolution explosion simulation of a 0.8 M$_\odot$ progenitor with initial mass fractions of 50$\%$ $^{12}$C, 48.9$\%$ $^{16}$O, 1$\%$ $^{22}$Ne and 0.1$\%$ $^{56}$Fe of Shen18.
The Shen18 explosion simulation used the MESA nuclear reaction network during the explosion simulation with a 41-isotope network, and $|\Delta$ln $T|_{max}$ was set to 0.04.
The explosion simulation was run out to 10 seconds.
We post-processed the density-temperature histories from these tracers using the MESA one zone burner as with the other tracer particles in the work.
To keep things consistent, we once again use the 205-isotope network, the initial mass fractions are set to 50$\% ^{12}$C, 48.6$\% ^{16}$O, and 1.4 $\%$ metals at solar mass fractions, and although the histories run out to 10 seconds, we stop at 7.5 seconds to compare results.

\begin{figure*}
	\centering\includegraphics[scale=0.85]{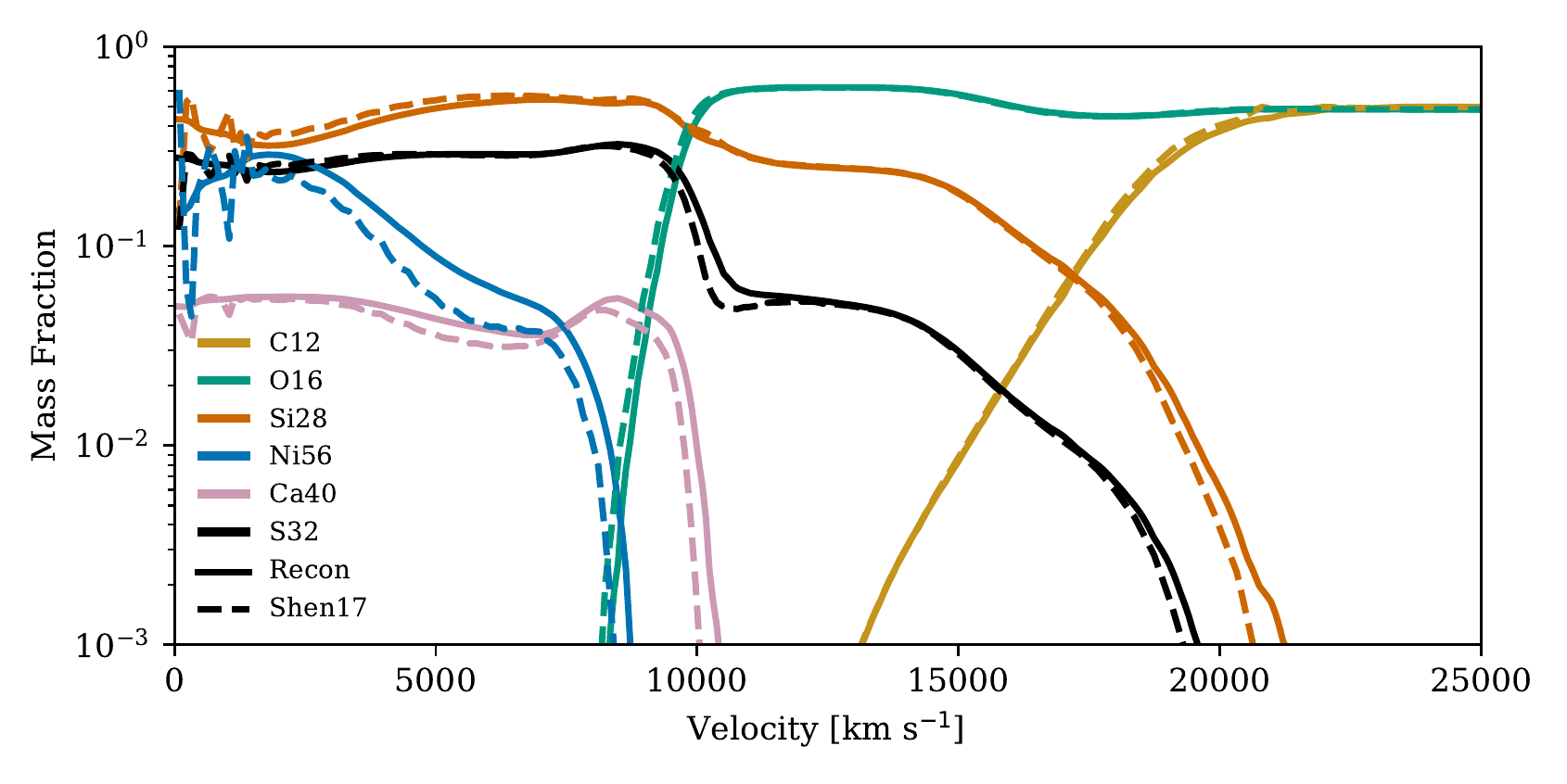}
	\caption{Mass fraction vs velocity profile for the results of the post-processing with detonation reconstruction (solid) and the results of the post-processed Shen18 results (dashed)}
	\label{fig:shen_vel_comapare}
\end{figure*}

The results of post-processing the Shen18 tracers compared to the detonation reconstruction results can be seen in figure~\ref{fig:shen_vel_comapare}.
Overall the mass fraction profile shapes of the mass fraction profiles produced by the two methods is similar, with peaks and troughs at similar velocities.
Throughout much of the ejecta the mass fractions of $^{28}$Si and, especially $^{32}$S agree fairly well.
The trough in $^{32}$S that occurs at $\sim$11000 km s$^{-1}$ is a bit deeper in the Shen18 results.
From 0 to $\sim$2000 km s$^{-1}$ the mass fraction profiles from the Shen18 results are noisy with $^{56}$Ni mass fractions oscillating from 0.05 to 0.5.
Similar noise can be seen in the $^{28}$Si, $^{32}$S, and $^{40}$Ca profiles.
From $\sim$2500 to $\sim$9000 km s$^{-1}$, the detonation reconstruction method consistently produces more $^{56}$Ni and $^{40}$Ca so once again the detonation reconstruction method is predicting more complete burning in these regions. 

\begin{figure*}
	\centering\includegraphics[scale=0.85]{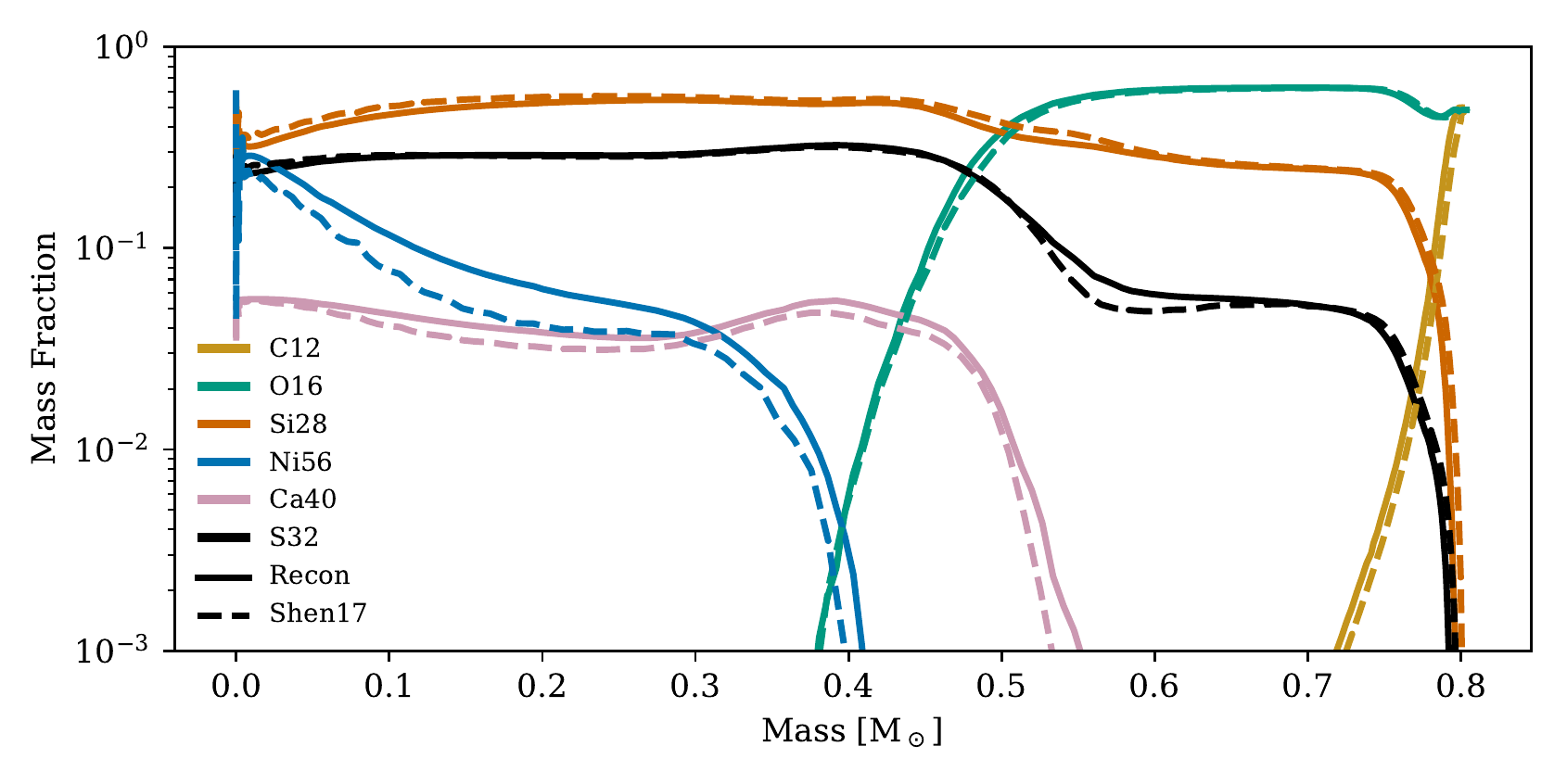}
	\caption{Mass fraction vs mass profile for the results of the post-processing with detonation reconstruction (solid) and the results of the post-processed Shen18 results (dashed)}
	\label{fig:shen_mass_comapare}
\end{figure*}

Figure~\ref{fig:shen_mass_comapare} shows the same results but in integrated mass coordinate. 
The noise seen in the Shen18 results at low velocity make up only a small part of the mass so it should have negligible effects on the overall yields.
It's clearer in this view that the detonation reconstruction method produces more $^{56}$Ni through a sizable fraction of the star, and that the burning is simply more complete in the detonation reconstruction case than in the Shen18 tracers.
Table~\ref{tab:shen_yields} gives a comparison between the total yields produced by the two methods. 
Overall, the detonation reconstruction method produces 6$\%$ less $^{28}$Si, 0.6$\%$ more $^{32}$S, 12$\%$ more $^{40}$Ca, and 34$\%$ more $^{56}$Ni than the post-processed Shen18 tracers.
The small amount of $^{56}$Ni produced in this 0.8 M$_\odot$ progenitor likely overemphasizes the fractional difference, and the $^{56}$Ni yields are expected to be closer on a fractional bases for higher mass progenitors. 

\begin{figure*}
	\centering\includegraphics[scale=1]{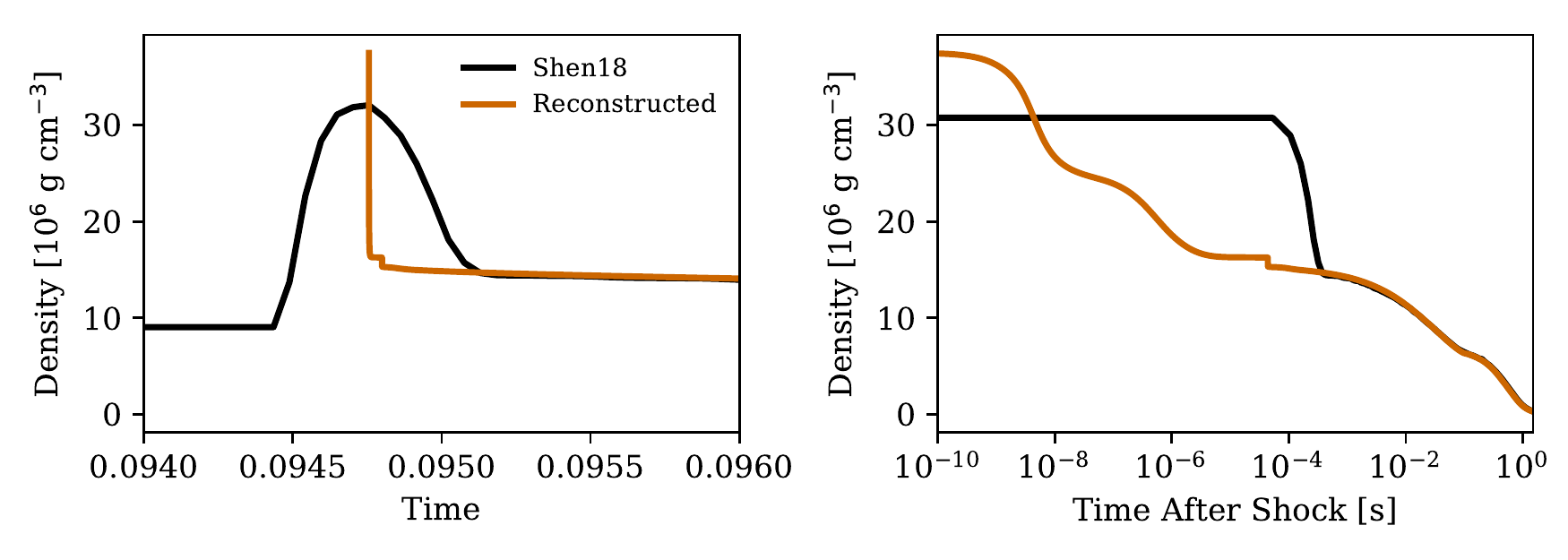}
	\centering\includegraphics[scale=1]{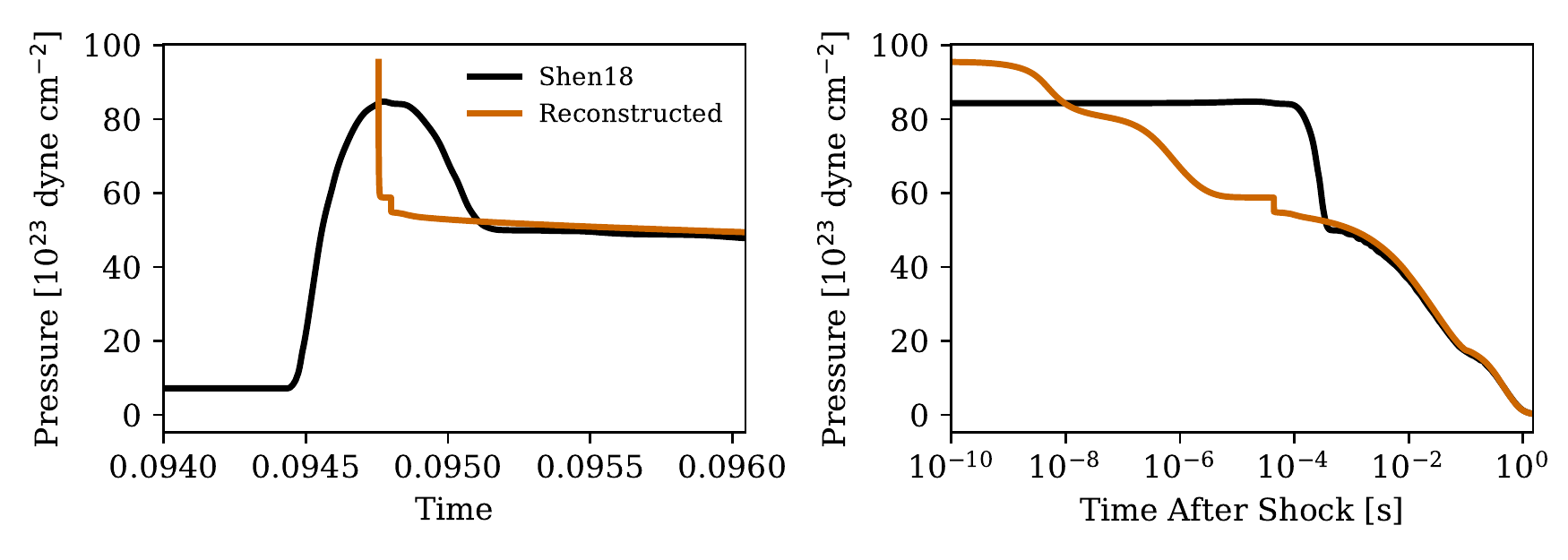}
	\centering\includegraphics[scale=1]{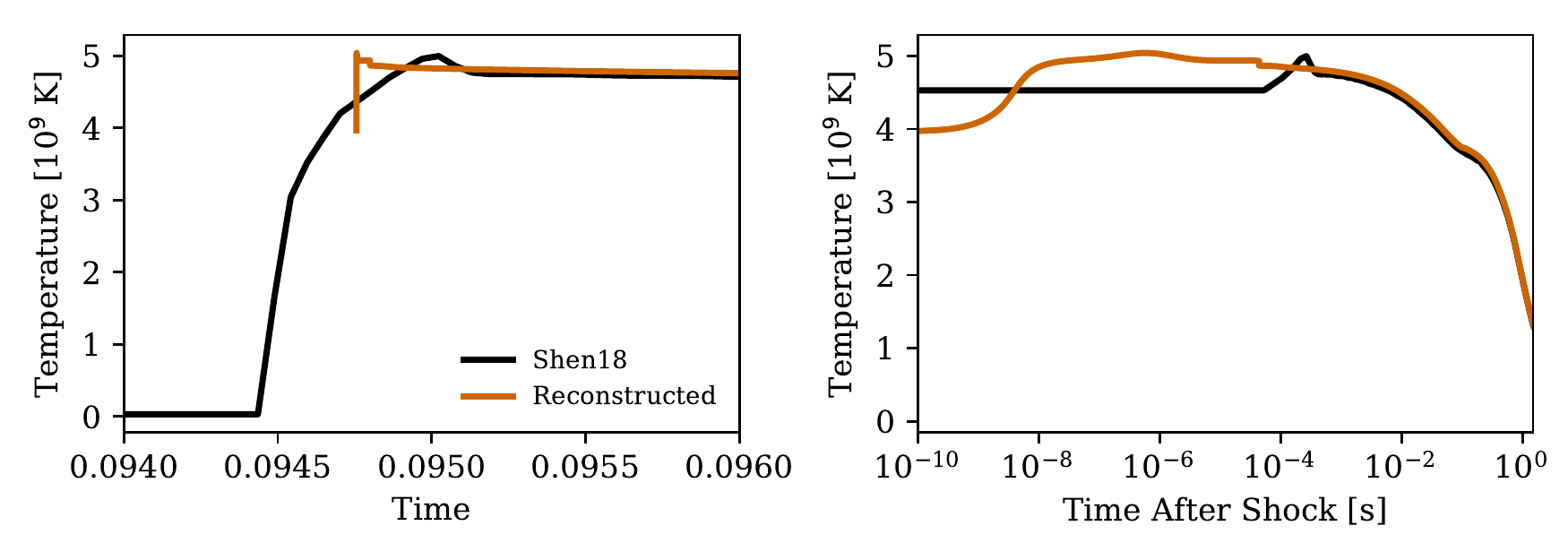}
	\caption{\label{fig:shen_comparison} Comparisons of density (top), pressure (middle), and temperature (bottom) as a function of time with the zero-point of the reconstructed track shifted to the time of the Shen18 density peak (left) and time after $\rho_{max}$ (right) for a reconstructed track (orange) and a track from Shen18 (black) at a pre-ignition density of 9.0$\times$10$^{6}$ g cm$^{-3}$. }
\end{figure*}

Why do these two methods that resolve the detonation structure differ by such large amount in the $^{56}$Ni yields?
To investigate, we take one track post-processed using the detonation reconstruction method and one track from the Shen18 results with pre-detonation densities of 9$\times$10$^{6}$ g cm$^{-3}$ and final mass coordinate of $\sim$0.03 M$_\odot$ where $^{56}$Ni should be produced and compare their density, pressure, temperature and mass fraction histories.
Figure~\ref{fig:shen_comparison} shows the density (top), pressure (middle), and temperature (bottom) histories for the Shen18 track (black) and reconstructed track (orange) as a function of time with the zero-point of the reconstructed track shifted to the time of the density peak (left) and the time after density peak (right).
The effect of the thickened detonation front is apparent; the density peak from the Shen18 track is broad in time, lasting several 10$^{-4}$ seconds.
Aside from the width of the density peak, the density does not reach the value obtained by the reconstructed detonation.
This points to the thickened detonation not being as strong as the reconstructed detonation.
Likewise, the pressure peak is wider than and does not reach the peak value obtained by the reconstructed detonation.
The temperature peak from the Shen18 track though wider in time, reaches a similar peak value as the reconstructed structure.
The rate-limiting has restored the separation between the pressure peak and temperature peak as intended. 

\begin{figure}
	\centering\includegraphics[width=\columnwidth]{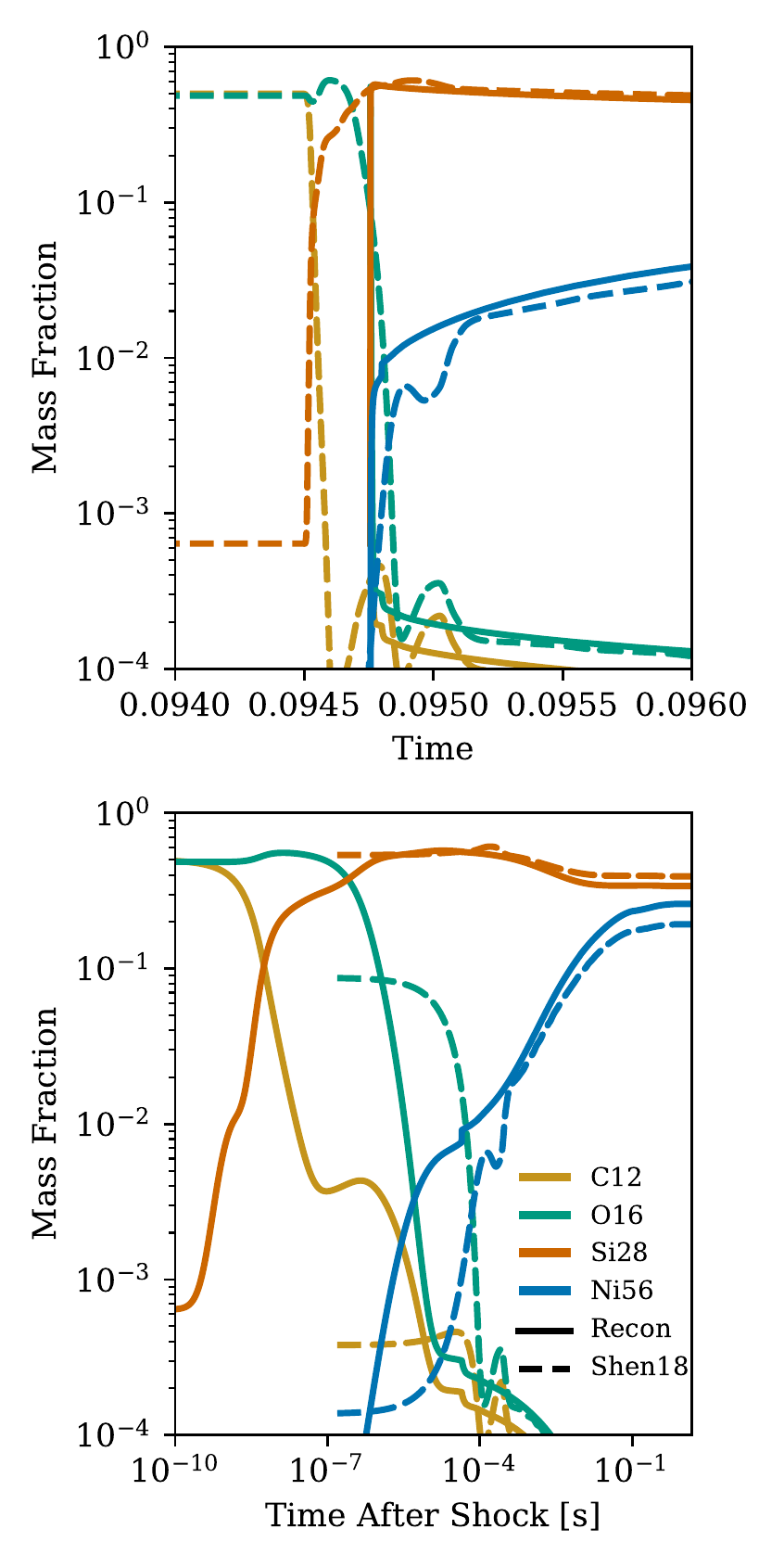}
	\caption{\label{fig:shen_abund}Mass fraction profiles in time with the zero-point of the reconstructed track shifted to the time of the Shen18 density peak (top) and time after $\rho_{max}$ (bottom) for a reconstructed track (solid) and a track from Shen18 (dashed) at a pre-ignition density of 9.0$\times$10$^{6}$ g cm$^{-3}$. The stronger reconstructed detonation results in more fully burned material. The mass fractions of the plotted species produced by an eigenvalue detonation at this density and curvature are nearly identical to those of the reconstructed results on this scale.}
\end{figure}

Figure~\ref{fig:shen_abund} shows the mass fractions of major isotopes produced by the Shen18 track (dashed) and detonation reconstruction (solid) during post-processing.
As expected from the pressure and density comparisons, the stronger reconstructed detonation results in material that is slightly more completely burned.
Both tracks produce similar mass fractions of $^{28}$Si, but the reconstructed track produces slightly more $^{56}$Ni. 

\begin{figure}
	\includegraphics[width=\columnwidth]{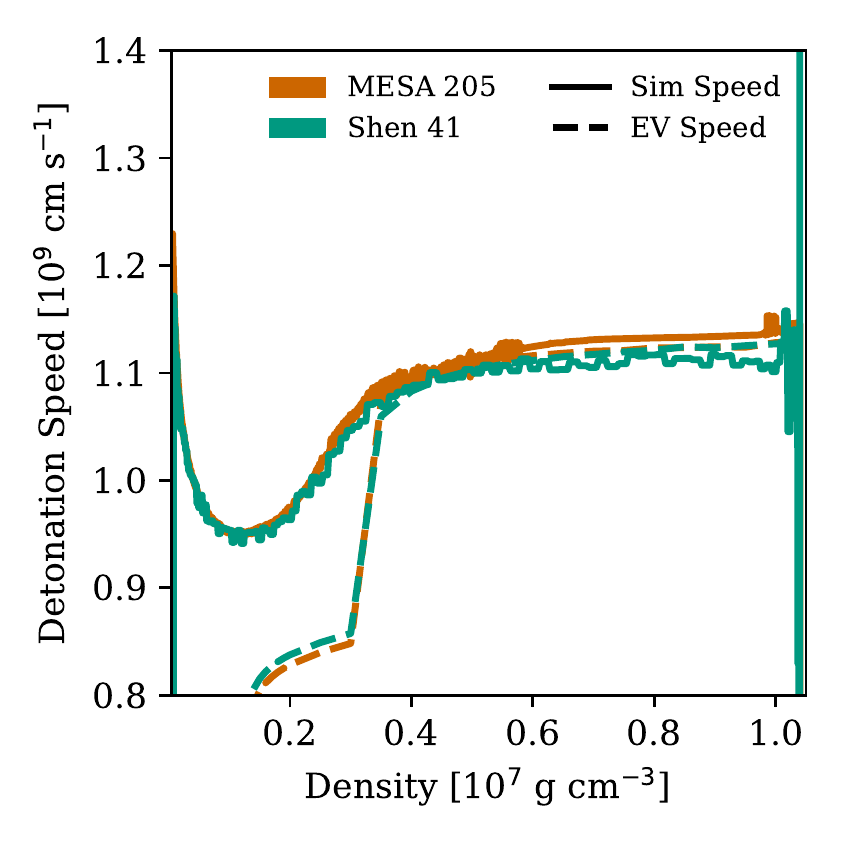}
	\caption{\label{shen_speed}Detonation speeds from explosion simulations (solid lines) using the MESA205 network (orange) and the Shen41 network with a burning limiter (green) compared to eigenvalue detonation speeds (dashed) with the MESA205 (orange) and Shen41 networks (green). The eigenvalue speed is similar for both networks. The detonation from the explosion simulation using the MESA205 network is faster than the Shen18 detonation through the majority of the star. Once both detonations reach the density gradient the speeds become comparable.}
\end{figure}

The explosion simulation that produced the tracers used in the post-processing comparison differed in two major ways from the explosion simulation used in the detonation reconstruction method: the thickened detonation front and the smaller nuclear reaction network.
The top and middle panels of figure~\ref{fig:shen_comparison} both appear to indicate that the detonation in the Shen18 calculation is weaker than the detonation in our explosion simulation.
In figure~\ref{shen_speed}, we calculate the eigenvalue speeds for a detonation using the 41-isotope network (dashed, green) and compare it to the eigenvalue speeds for a detonation powered by the 205-isotope network (dashed, orange).
The eigenvalue speeds are nearly identical except at the lowest densities so it doesn't appear the network choice is driving the differences.
The mass fractions of the plotted species produced by an eigenvalue detonation at this density and curvature are nearly identical to those of the reconstructed results on this scale.
The measured detonation speed from the Shen18 explosion simulation is also shown in this figure.
Where our detonation is overdriven through much of the star, the Shen18 detonation travels at speeds lower than the eigenvalue speed.
The detonation is indeed weaker, and this is what is leading to the differences in produced yields of $^{40}$Ca and $^{56}$Ni. 
This provides good confidence that these two methods bound the actual yield from above and below.
\begin{deluxetable}{c|cc}
	\tablewidth{\columnwidth}
	\tablecaption{Yield in M$_\odot$ Comparison: Post-processing with Detonation Reconstruction vs Shen18 + Direct Post-Processing \label{tab:shen_yields}}
	\tablehead{%
		\colhead{Iso} & \colhead{Reconstruction} & \colhead{Shen18 + Post-Processing}
	}
	\startdata
	$^{28}$Si & 0.319 & 0.340 \\
	\hline
	$^{32}$S\phm{.}\Tstrut & 0.161 & 0.160 \\
	\hline
	$^{40}$Ca\Tstrut & 0.0220 & 0.0196 \\
	\hline
	$^{56}$Ni\Tstrut & 0.0350 & 0.0262
	\enddata
\end{deluxetable}

\section{Quantifying Reconstruction Uncertainty}
\label{sec:uncertainty}
A source of uncertainty in the detonation reconstruction method is the detonation speed that is used in the calculation of the steady-state overdriven detonation used for part of our approximate quasi-steady-state structure.
Presently, we have chosen to use the detonation speed measured from the explosion simulation.
The forward propagation of the detonation front is powered by the fusion reactions occurring behind the front.
The speed of the eigenvalue detonation is determined by the total amount of energy released in the reaction zone up to the sonic point in the case of an eigenvalue detonation.

The energy release of the detonation, Q, can be determined from the  difference between the initial, pre-detonation mass fractions and final mass fractions at the sonic point for the eigenvalue detonation.
An overdriven detonation, such as the detonation in our simulations, does not have a sonic point.
To calculate the energy release from the overdriven detonation, we evaluate the mass fractions at the point of minimum pressure behind the shock which is analogous to the sonic point of the eigenvalue detonation.
The unresolved nature of the detonation causes the nucleosynthetic products in the explosion simulation to be suspect.
However, if the overall energy released by the detonation is similar to that of a resolved detonation, then using the detonation speed from the explosion simulation is an appropriate first choice.
To determine if the energy release of the explosion simulation is reasonable, we compare the energy release of the detonation from the explosion simulation to the energy release of the steady-state detonation at the same density and temperature with the same radius of curvature and detonation speed.

\begin{figure}
\includegraphics{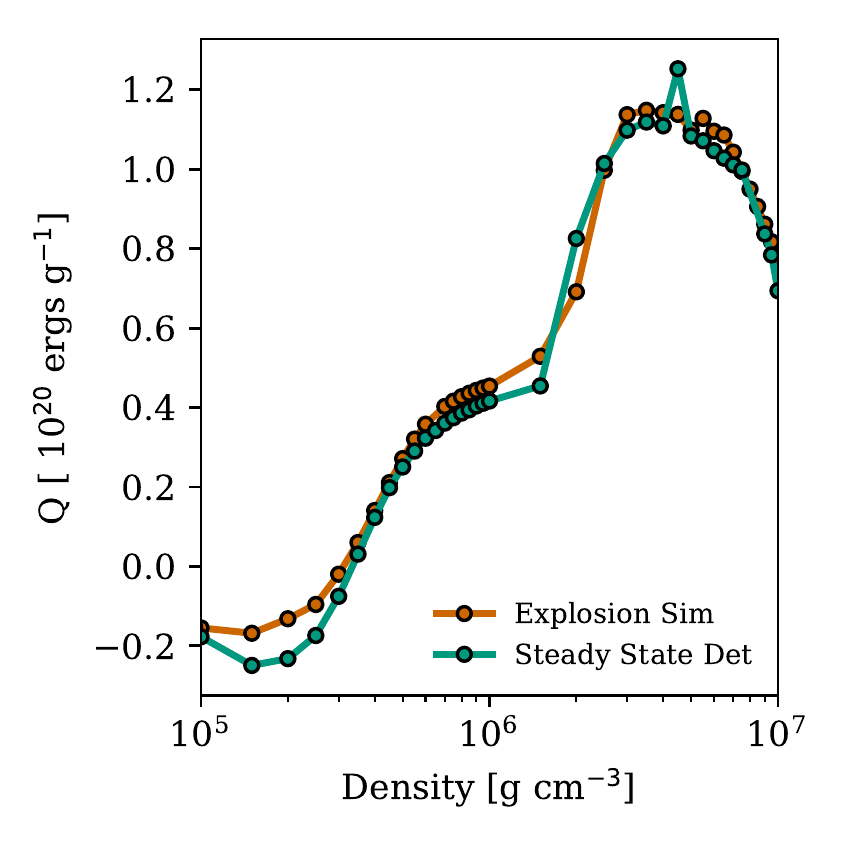}
\caption{Q values calculated using the mass fractions of the nucleosynthetic products from the 0.0625~km resolution explosion simulation (orange) and steady-state overdriven detonation at the speed observed in the simulation (green). Negative Q values arise from the production of alpha particles, reducing the amount of energy available to power the propagation of the detonation}
\label{fig:q_values}
\end{figure}

Figure~\ref{fig:q_values} shows the calculated Q values as a function of density.
As a consequence of the tracer particles only carrying the mass fractions of $^{12}$C, $^{16}$O, $^{28}$Si, $^{32}$S, $^{40}$Ca, and $^{56}$Ni, the Q values were calculated based only on the initial and final mass fractions of these isotopes.
Though this not account for all the present mass, these six species make up $\sim$90\% of fuel and burning products, and for the purposes of this check, are sufficient.
Over the majority of the shown densities, the Q values from the explosion simulation are higher than the Q values from the steady-state detonation.
A notable deviation occurs at 4$\times$10$^{6}$~g~cm$^{-3}$, where the steady state detonation at the observed speed has a larger Q value than the explosion simulation case.
The detonation speed in this density range (figure~\ref{det_speed_fig}) is oscillating and is in the region where the eigenvalue detonation transitions to the region of weaker, slower detonation, only producing oxygen and silicon-group material (figure~\ref{curv_fig}).
It is possible that the detonation in the explosion simulation may be attempting to transition to the lower speed branch as well, and may be limited by the resolution.
Another area where the Q values from the two calculations differ by a wider margin is in the lower density region between 1$\times$10$^{5}$~g~cm$^{-3}$ and 4$\times$10$^{5}$~g~cm$^{-3}$.
Here, the detonation is accelerating due to the density gradient in the outer radii of the white dwarf.
The strengthening of the detonation is a dynamic problem and modeling its behavior with a steady-state detonation may not be appropriate.

The similarity in Q value obtained from the mass fractions in the simulation itself and for a steady-state detonation at the observed speed lend support to the conclusions made from the comparison of the detonation speed to the eigenvalue speed.
At densities above a few $10^6$ g cm$^{-3}$, the small additional energy release present in the simulation is likely leading to a modest overdriving of the detonation in this region.
At the lower densities, the similarity of the Q values indicates that the shock strengthening due to the density gradient is dominant over the driving due to energy release.

Although we do not explore it here, the differences in Q value shown in figure~\ref{fig:q_values} could be used to quantify the uncertainty in the detonation speed.
Since the speed is a parameter in the reconstruction method, it could be varied over the indicated range to better quantify the remaining systematic uncertainty. In any such consideration, we would also recommend variation of the location of the pasting point as the other major parameter in the reconstruction method.

\section{Discussion and Conclusion}
In this work, we have investigated the behavior of detonations in the explosions of  white dwarfs.
By using a large nuclear reaction network in both the explosion simulation and particle post-processing, we have reduced the uncertainty that can be introduced by running explosion simulations with smaller networks and then post-processing with larger networks.
However, due to the unresolved nature of the detonation, the nucleosynthetic yields produced by the explosion simulation are not sufficiently accurate despite the use of a large nuclear reaction network.
To mitigate this issue, we have developed a new method of post-processing in which the unresolved structure of the detonation is replaced with results of a steady-state detonation for each temperature-density history.

In the process of analyzing this new method and comparing in detail to both steady-state detonations and other methods, we have demonstrated that both the curvature of the detonation front and the strengthening that the shock experiences going down the density gradient present in the star are essential for determining the yields driven by carbon detonations in Type Ia supernovae. 
Our reconstruction method is able to account for both of these effects in a way that allows, for the first time, yields to be computed with fully-time-resolved thermodynamic histories.
We show that our simulations display detonations that are slightly stronger than the expected steady-state solutions at densities above a few $\times$10$^{6}$ g cm$^{-3}$, and at lower densities driven well above the steady-state speed by shock strengthening due to the density gradient present in the star.
Figures~\ref{fig:shen_vel_comapare}~and~\ref{fig:shen_mass_comapare} comprise the principal results of this work.
By comparing to other methods and the steady-state speeds, we estimate that remaining systematic uncertainty is well below 10\%, but that it will require improved subgrid modeling of reactions in the simulation before yields can be verified to better than 1\% uncertainty.
Note that this does not include uncertainty in nuclear reaction rates.

The reconstruction method still faces several challenges, particularly for application in multi-dimensional simulations.
Two of the input parameters, the radius of curvature of the detonation front and detonation speed, must be measured in some way before reconstruction can be done.
For one dimensional, centrally ignited detonations, the first is trivial.
The radius of curvature in that case is simply the radial location of the detonation front.
However, when moving into multiple dimensional calculations, the detonation may no longer be centrally ignited, and the radius of curvature will vary with position along the detonation front.
We currently do not have the capability of measuring this.
Determining the correct detonation speed is more problematic.
Here we have used the measured shock speed from the explosion simulation as our detonation speed, but once again, this is only possible due to the calculation only being one dimensional.
Multidimensional phenomena such as the formation of cellular structures create problems for this technique.
Also, the results of section~\ref{sec:uncertainty} show that the energy release from the detonation does not match the energy release of a steady state detonation.
These problems show the need of the creation of a sub-grid model that properly captures all of the relevant detonation physics.

One piece of critical physics that current sub-grid detonation models lack is shock strengthening due to the density gradient of the white dwarf.
Contrary to the results of \cite{Dunkley_et_al_2013}, the detonations in our calculations are not extinguished by the effects of curvature.
As shown in figure~\ref{det_speed_fig}, the detonation not only remains active, but accelerates.
Shocks strengthening due to a density gradient is well known phenomena, but it has not been taken into account in models of SNe Ia.

The dependence of the local detonation characteristics on the local environment, as captured by the subgrid model, is eminently important for high-accuracy yields needed to test proposed models of SNe Ia.
Use of a subgrid model that depends only on the local density and is globally calibrated, as is done in \cite{Fink_2010} and related work is insufficient.
Even methods proposed in literature on terrestrial detonations \citep{bdzil_review} that consider the local density and curvature are insufficient due to the crucial importance of the density gradient.
In our simulations, the detonation is forced to travel directly down the density gradient, resulting in maximum strengthening.
However, in the case of double detonation SNe Ia, the detonation may travel directly from the outer helium shell into the white dwarf.
When this occurs, the detonation will be traveling at some angle with respect to the density gradient, strengthening it or weakening it.
Likewise, if the detonation of the white dwarf is triggered by compressive shocks from the helium shell detonation, the detonation may begin off-center, and different regions of the detonation front will experience varying effects from the density gradient.
Also, in multi-dimensional calculations, detonations take on a complex cellular structure \citep[see][for more details]{papatheodore}, and it is not clear how the density gradient will affect this.
\\\\
B.J.M., D.M.T., and K.J.S.\ received support from NASA through the Astrophysics Theory Program (NNX17AG28G). Portions of this work were supported by the United States Department of Energy under an Early Career Award (DOE grant no. SC0010263) and by the Research Corporation For Science Advancement under a Cottrell Scholar Award.

\software{FLASH \cite{fryxell_2000_aa}, eZND \citep{Moore_et_al_2013}, MESA \citep{paxton_2011,paxton_2013,paxton_2015,paxton_2018}}

\bibliography{master,timmes_master,townsley_master,detonations}

\end{document}